\documentclass[12pt,a4paper]{article}

\usepackage{epsfig}
\usepackage{amsmath}
\usepackage{amssymb}
\usepackage{verbatim}
\usepackage{psfrag}
\usepackage{bm}
\usepackage{dsfont}
\usepackage[footnotesize]{caption}

\newcommand{\dif}{\,\mbox{d}}

\newcommand{\ophi}{\bar{\phi}}
\newcommand{\oH}{\bar{H}}
\newcommand{\oR}{\bar{R}}
\newcommand{\oL}{\bar{L}}
\newcommand{\oF}{\bar{F}}
\newcommand{\ou}{\bar{u}}
\newcommand{\od}{\bar{d}}
\newcommand{\oel}{\bar{e}}
\newcommand{\on}{\bar{\nu}}
\newcommand{\tnu}{\tilde{\nu}}

\newcommand{\sixteen}{\mathbf{16}}
\newcommand{\csixteen}{\mathbf{\overline{16}}}

\newcommand{\euler}{\text{e}}
\newcommand{\imag}{\text{i}}

\newcommand{\SOten}{SO(10)}

\newcommand{\Zten}{\mathds{Z}_{10}}
\newcommand{\Ztwo}{\mathds{Z}_{2}}
\newcommand{\rep}{\mathbf}
\newcommand{\MP}{M_{\text{P}}}

\newcommand{\PS}{SU(4)_{C}\times SU(2)_{L}\times SU(2)_{R}}
\newcommand{\SM}
{SU(3)_{C}\times SU(2)_{L}\times U(1)_{Y}}
\newcommand{\GPS}{G_{\text{PS}}}
\newcommand{\GSM}{G_{\text{SM}}}

\begin{document}

\begin{flushright}
\tt{UG-FT/266/10, CAFPE/136/10}\\
\tt{MPP-2010-30, SHEP-10-09}
\end{flushright}

\vskip 0.05cm

\begin{center}
{\Large\bf Gauge Non-Singlet Inflation in SUSY GUTs}

\vskip 1cm

Stefan~Antusch,$^1$ Mar~Bastero-Gil,$^2$ Jochen~P.~Baumann,$^1$ Koushik~Dutta,$^1$ \\ Steve~F.~King$^3$ and Philipp~M.~Kostka$^1$\\[3mm]
{$^1$\it{
Max-Planck-Institut f\"ur Physik (Werner-Heisenberg-Institut),\\
F\"ohringer Ring 6,
80805 M\"unchen, Germany}\\[2mm]
$^2$\it{Departamento de Fisica Teorica y del Cosmos and
Centro Andaluz de Fisica de Particulas Elementales,
Universidad de Granada, 18071 Granada, Spain\\[2mm]
$^3$\it{School of Physics and Astronomy, University of Southampton,\\
Southampton, SO17 1BJ, United Kingdom}} }
\end{center}

\vskip 0.75cm

\begin{abstract}
We explore the novel possibility that the inflaton responsible for
cosmological inflation is a gauge {\it non}-singlet
in supersymmetric (SUSY) Grand Unified Theories (GUTs).
For definiteness we consider SUSY hybrid inflation
where we show that the scalar components of gauge non-singlet superfields,
together with fields in conjugate representations, may form a D-flat direction suitable for inflation.
We apply these ideas to SUSY models with an Abelian gauge group, a Pati-Salam gauge group and finally
Grand Unified Theories based on $SO(10)$ where the scalar components of the
matter superfields in the $\sixteen$s may combine with a single $\csixteen$ to form the inflaton, with
the right-handed sneutrino direction providing a possible viable trajectory for inflation.
Assuming sneutrino inflation,
we calculate the one-loop Coleman-Weinberg corrections 
and the two-loop corrections from gauge interactions giving rise to the ``gauge $\eta$-problem'' and show that both corrections
do not spoil inflation, and the monopole problem can be resolved.
The usual $\eta$-problem arising from supergravity may also be resolved
using a Heisenberg symmetry.
\end{abstract}

\newpage
\tableofcontents
\newpage

\section{Introduction}
The inflationary paradigm remains extremely successful in solving the
horizon and flatness problems of the standard Big Bang cosmology,
and at the same time in explaining the origin of structure of the
observable Universe~\cite{Guth:1980zm,Liddle:2000cg}.
Several schemes for inflation have been proposed including
chaotic inflation~\cite{Linde:1983gd}, which
predicts large tensor perturbations~\cite{Lyth:1996im},
in contrast to hybrid inflation~\cite{Linde:1991km, Linde:1993cn, Linde:1997sj, Jeannerot:1997is, BasteroGil:1999fz}
which predicts small ones. The main advantage of hybrid inflation is that, since it involves small
field values below the Planck scale, it allows a small
field expansion of the K\"ahler potential in the
effective supergravity (SUGRA) theory, facilitating the connection
with effective low energy particle physics models such as SUSY
extensions of the Standard Model (SM) and GUTs~\cite{Dvali:1994ms}.

A long standing question in inflation models is: Who is the inflaton?
We are still far from answering this question. Indeed it is still unclear
whether the inflaton, the (presumed) scalar field responsible for inflation,
should originate from the observable (matter) sector or the hidden (e.g.~moduli) sector
of the theory. However the connection between inflation and particle physics is rather difficult
to achieve in the observable sector due to the lack of understanding of physics beyond the
Standard Model (SM) and in the hidden sector due to the lack of understanding of
the string vacuum. However over the past dozen years there has been a revolution in
particle physics due to the experimental discovery of neutrino mass and mixing~\cite{King:2003jb},
and this improves the prospects for finding the inflaton in the observable sector.
Indeed, if the SM is extended to include the seesaw mechanism~\cite{seesaw}
and SUSY~\cite{Chung:2003fi}, the right-handed sneutrinos, the superpartners of the
right-handed neutrinos, become excellent inflaton candidates.
Motivated by such considerations, the possibility of chaotic (large field) 
inflation with a sneutrino inflaton~\cite{Murayama:1992ua} was
revisited~\cite{Ellis:2003sq}. Subsequently three of us with Shafi suggested that one (or more) of the
singlet sneutrinos could be the inflaton of hybrid inflation~\cite{Antusch:2004hd}.

Despite the unknown identity of the inflaton, conventional wisdom dictates that
it must be a gauge singlet since otherwise
radiative corrections would spoil the required flatness of the inflaton potential.
For example in SUSY models scalar components of gauge non-singlet superfields have quartic terms in their potential,
due to the D-terms, leading to violations of the slow-roll conditions which
are inconsistent with recent observations by WMAP. In addition, gauge non-singlet
inflatons would be subject to one-loop Coleman Weinberg corrections from loops with gauge fields
which could easily lead to
large radiative corrections that induce an unacceptably large slope of the inflaton potential.
Furthermore a charged inflaton is in general also subject to two-loop corrections to its mass 
which can easily be larger than the Hubble scale~\cite{Dvali:1995fb}.
Such a contribution is in principle large enough to spoil inflation for any gauge non-singlet scalar field,
leading to a sort of ``gauge $\eta$-problem''.

In this paper we shall argue that, contrary to conventional wisdom, the inflaton may in fact
be a gauge {\it non}-singlet (GNS).
For definiteness we shall confine ourselves here to examples of
SUSY hybrid inflation \footnote{We note that GNS inflation may be applied to
other types of inflation other than SUSY hybrid inflation.} and show that the scalar components of gauge non-singlet superfields,
together with fields in conjugate representations, may form a D-flat direction suitable for inflation. Along this D-flat trajectory the usual F-term contributes the large vacuum energy. 
We apply these ideas first to a simple Abelian gauge group $G=U(1)$, then to a realistic SUSY Pati-Salam model,
then to $SO(10)$ SUSY GUTs, where the scalar components of the
matter superfields in the $\sixteen$s may combine with a single $\csixteen$ to form the inflaton, with
the right-handed sneutrino direction providing a possible viable trajectory for inflation.

We emphasize that, in sneutrino inflation models, the right-handed sneutrino has previously been
taken to be a gauge singlet, as for example in SUSY GUTs based on $SU(5)$ rather than $SO(10)$.
However, one of the attractive features of $SO(10)$ SUSY GUTs
is that it {\it predicts} right-handed neutrinos
which carry a charge under a gauged $B-L$ symmetry. The right-handed sneutrinos of
SUSY $SO(10)$, being charged under a gauged $B-L$ symmetry,
have not previously been considered as suitable inflaton candidates, but here they may be.
Indeed, assuming the sneutrino inflationary trajectory, 
we calculate the one-loop Coleman-Weinberg corrections
and the two-loop corrections usually giving rise to the ``gauge $\eta$-problem'' and show that both corrections do not spoil inflation.
In addition we show that the monopole problem~\cite{Kibble:1976sj} 
of $SO(10)$ GUTs can be resolved.
We shall also show that the usual $\eta$-problem arising from SUGRA~\cite{Copeland:1994vg} may be resolved
using a Heisenberg symmetry~\cite{Binetruy: 1987}
with stabilized modulus~\cite{Antusch:2008pn}.

The layout of the remainder of the paper is as follows.
In Section~2 we introduce the idea of SUSY hybrid inflation
with a GNS inflaton, focusing on the example of an Abelian
gauge group $G=U(1)$.
In Section~3 we discuss a realistic model of this kind
based on the SUSY Pati-Salam gauge group, specializing to the
case of the right-handed sneutrino inflationary trajectory.
In Section~4 we embed the preceding Pati-Salam model into
$SO(10)$ SUSY GUTs. Section~5 confronts the issues associated
with radiative corrections for a GNS inflaton at one and two loops.
Section~6 shows how the $\eta$\,-\,problem in SUGRA may be resolved in this
class of models using a Heisenberg symmetry with stabilized modulus.
Section~7 summarizes and concludes the paper.


\section{SUSY Hybrid Inflation with a GNS Inflaton}\label{Toy Model and Basic Ideas}

SUSY hybrid inflation is typically based on the superpotential
\cite{Dvali:1994ms}
\begin{equation}\label{susyhybrid}
W_0=\kappa\,S\left(H\oH-M^2\right)
\end{equation}
where the superfield $S$ is a singlet under some gauge group $G$, while
the superfields $H$ and $\oH$ reside in conjugate
representations (reps) of $G$. The F-term of $S$ provides the
vacuum energy to drive inflation, the scalar component of
the singlet $S$ is identified as the slowly rolling inflaton, and the scalar components
of $H$ and $\oH$ are waterfall fields which take zero values during inflation
but are switched on when the inflaton reaches some critical value, ending inflation
and breaking the gauge group $G$ at their global minimum
$\langle H \rangle = \langle \oH \rangle = M$. Typically $G$ is identified as a GUT group
and $H$, $\oH$ are the Higgs which break that group~\cite{Dvali:1994ms}.

Consider the following simple extension of the superpotential in Eq.~\eqref{susyhybrid},
\begin{equation}\label{toymodel}
W=W_0+\frac{\zeta}{\Lambda}\left(\phi\,\ophi\right)\left(H\oH\right)
\end{equation}
where we have included an additional pair of GNS superfields $\phi$ and $\ophi$ in conjugate reps of $G$ which couple to the Higgs superfields via a non-renormalizable coupling controlled by a dimensionless coupling constant $\zeta$ and a scale $\Lambda$.\footnote{For illustrative purposes in this section we only consider the single operator contraction 
shown even though other distinct operators with different contractions are expected. A fully realistic model of this type will be presented in the next section.}
At first glance, we might expect the presence of the effective operator in Eq.~\eqref{toymodel}, that we have added to the superpotential $W_0$ in Eq.~\eqref{susyhybrid}, to not perturb the usual SUSY hybrid inflation scenario described above. However its presence allows the new possibility that inflation is realized via slowly rolling scalar fields contained in the superfields $\phi$ and $\ophi$ with the singlet field $S$ staying fixed at zero during (and after) inflation. In a SUGRA framework, non-canonical terms for $S$ in the K\"ahler potential can readily provide a large mass for $S$ such that it quickly settles at $S=0$. On the other hand, large SUGRA mass contributions can be avoided for $\phi$ and $\ophi$ using a Heisenberg symmetry~\cite{Antusch:2008pn} as will be briefly discussed in section~\ref{Generalization to Supergravity}.

While the singlet $S$ field is held at a zero value by SUGRA corrections,
the scalar components of $\phi$, $\ophi$, having no such SUGRA corrections,
are free to take non-zero values during the inflationary epoch.
The non-zero $\phi$, $\ophi$ field values provide
positive mass squared contributions to all components of the
waterfall fields $H$ and $\oH$ during inflation, 
thus stabilizing them at zero by the F-term potential from the
second term in Eq.~\eqref{toymodel}.
As in standard SUSY hybrid inflation, the F-term of $S$, arising from $W_0$ in Eq.~\eqref{susyhybrid}, yields
a large vacuum energy density $V_0=\kappa^2 M^4$ which drives inflation and breaks SUSY.
Since $\phi$, $\ophi$ are the only fields which are allowed to take non-zero values during inflation,
they may be identified as inflaton(s) provided that their potential is sufficiently flat.
Since both $\phi$ and $\ophi$ carry gauge charges under $G$, their vacuum expectation values (VEVs)
break $G$ already during inflation, thus, although $\phi$ and $\ophi$ are GNS fields under the original gauge group
$G$, they are clearly gauge singlets under the surviving subgroup of $G'\subset G$ respected by inflation.
This trivial observation will help to protect the $\phi$ and $\ophi$
masses against large radiative corrections, as we shall
see later. Another key feature is that the quartic term in the $\phi$ and $\ophi$ potential arising from
D-term gauge interactions is
avoided in a D-flat valley in which the conjugate fields $\phi$ and $\ophi$ take equal VEVs.

Let us assume that the potential of $\phi$ and $\ophi$ is sufficiently flat to enable
them to be slowly rolling inflaton(s), and
that the dominant contribution to the slope of the inflaton potential arises from
quantum corrections due to SUSY breaking which make $\phi$ and $\ophi$ slowly roll towards zero.
Then the waterfall mechanism which ends inflation works in a familiar way, as follows.
Once a critical value of $\phi$ and $\ophi$ is reached, the negative mass squared contributions
to the scalar components of $H$ and $\oH$ (from $W_0$ in Eqs.~\eqref{susyhybrid},\eqref{toymodel})
dominate, destabilizing them to fall towards their true vacuum.
In this phase transition, the breaking of $G$ is basically ``taken over'' by the
Higgs VEVs $\langle\oH^*\rangle=\langle H\rangle \sim M$ and at the same time
inflation ends due to a violation of the slow-roll conditions.
The vacuum energy is approximately cancelled by the Higgs VEVs and SUSY is
approximately restored at the global minimum.

\subsection{Explicit Example with $G=U(1)$}
\begin{figure}[!h]
\psfrag{H}{\hspace{-1.5cm}\footnotesize$\text{Re}(\oH+H),\,\text{Im}(\oH-H)$}
\psfrag{N}{\footnotesize$|\phi|$}
\psfrag{V}{\footnotesize$V(\phi,H)$}
\center
\includegraphics[scale=0.8]{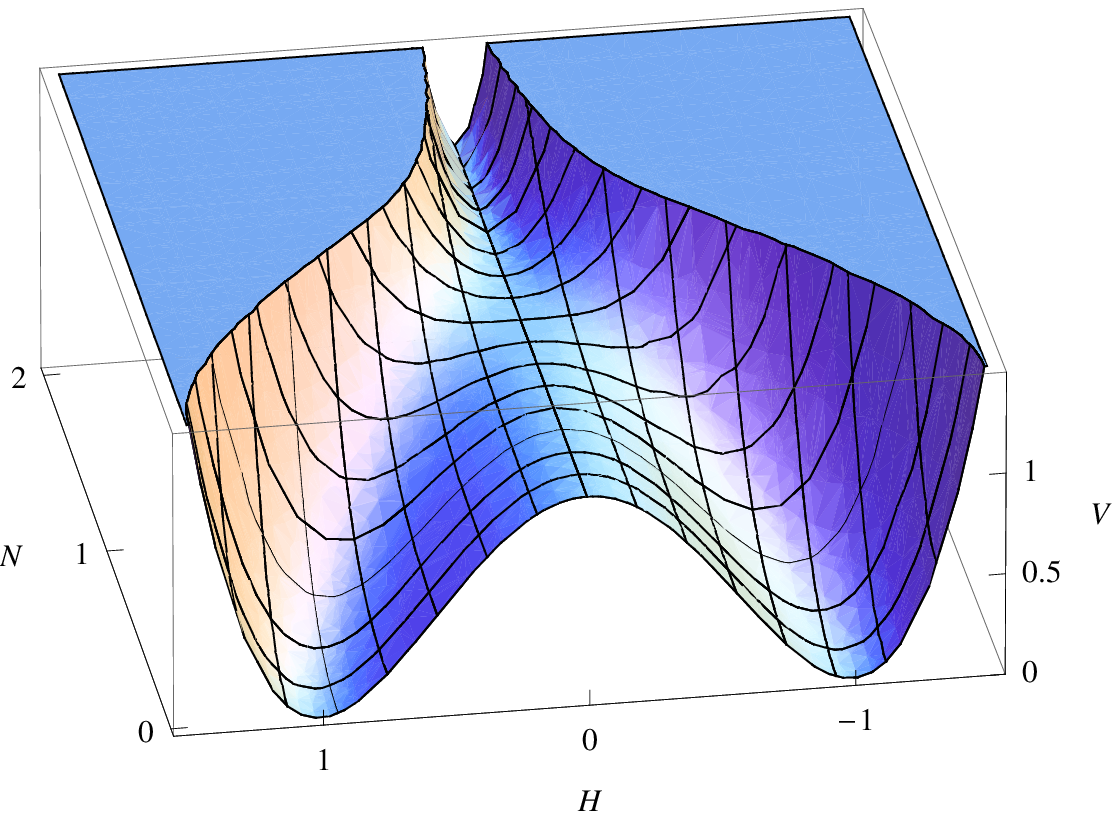}\\ \vspace{0.5cm}
\includegraphics[scale=0.8]{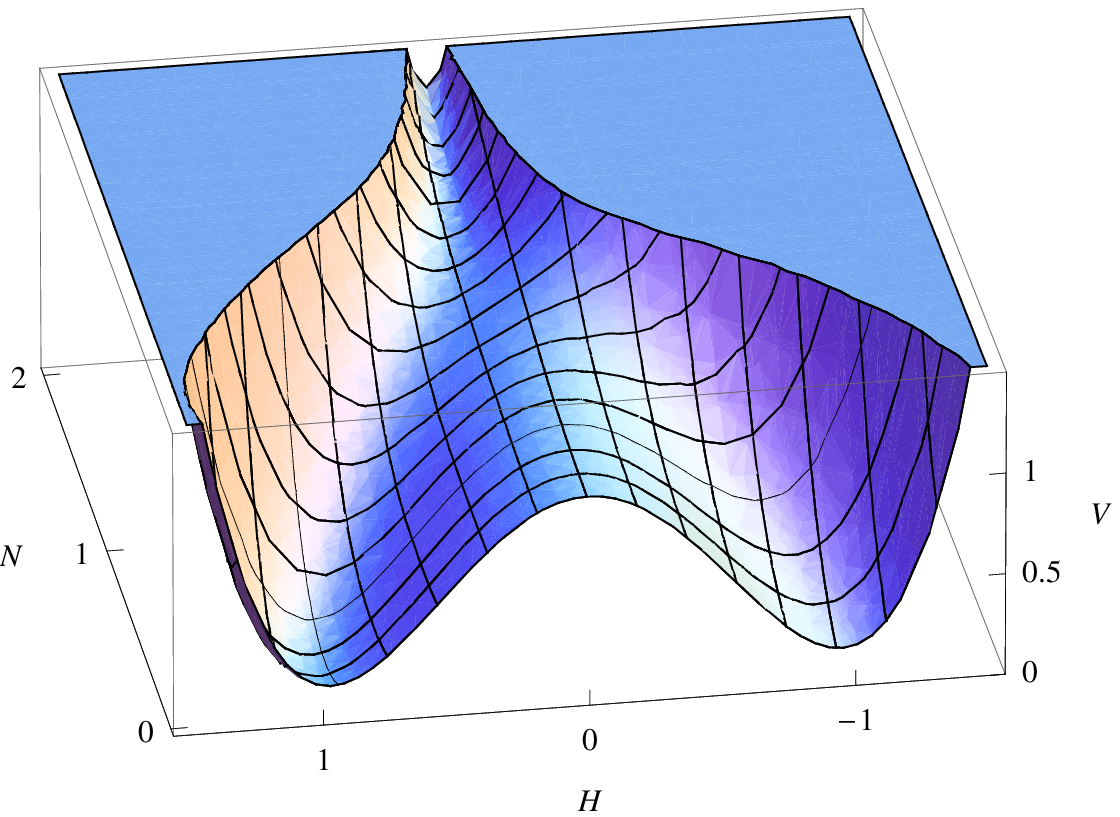}
\caption{\label{hybrid}Plot of the F-term hybrid inflation potential
in the D-flat valleys $\phi=\ophi^*$, $H=\oH^*$, without
deformations by higher-dimensional effective operators (upper plot).
The lower plot displays the deformed potential where an effective superpotential
term $\delta\,(H\ophi)$ has been switched on.
This term gives rise to a slope at $H=\oH=0$ that forces the field 
into the global minimum at positive $M$.}
\end{figure}

Let us now explicitly calculate the full global SUSY potential
for the model in Eq.~\eqref{toymodel},
assuming an Abelian gauge group $G=U(1)$.
Any SUSY gauge theory gives rise to a scalar potential
\begin{equation}
V=V_F+V_D=F^{*}_i F^i+\frac{1}{2}\,D^a D^a\,.
\end{equation}
For $G=U(1)$ and equal charge for $\phi$ and $H$ we find $D=-g\left(|\phi|^2-|\ophi|^2+|H|^2-|\oH|^2\right)$,
where the index $a$ has disappeared because a $U(1)$ has only one generator and
$g$ is the gauge coupling constant.
Thus we obtain a D-term contribution (setting a possible Fayet-Iliopoulos term to zero)
\begin{equation}\label{Dtermpotential}
V_D=\frac{g^2}{2}\left(|\phi|^2-|\ophi|^2+|H|^2-|\oH|^2\right)^2\,,
\end{equation}
which in the inflationary trajectory $\langle H\rangle=\langle\oH^*\rangle=0$ obviously has a
D-flat direction ${\langle\phi\rangle}=\langle\ophi^*\rangle$.
Under the assumption that the D-term potential Eq.~\eqref{Dtermpotential}
has already stabilized the fields in the D-flat valley, the remaining
potential is generated from the F-term part
\begin{equation}\label{Ftermpotential}
\begin{split}
V_F=&\left|\kappa\left(H\oH-M^2\right)\right|^2
+\,\left|\frac{\zeta}{\Lambda}\,\ophi\,(H\oH)\right|^2
+\,\left|\frac{\zeta}{\Lambda}\,\phi\,(H\oH)\right|^2\\
&+\,\left|\kappa \,S\,\oH+\frac{\zeta}{\Lambda}\,(\phi\,\ophi)\,\oH\right|^2
+\,\left|\kappa \,S\,H+\frac{\zeta}{\Lambda}\,(\phi\,\ophi)\,H\right|^2\,,
\end{split}
\end{equation}
which can be calculated with the equations of motion ${F^*}^i=-\partial W/\partial \phi_i$.
Plugging the D-flatness condition ${\langle\phi\rangle}=\langle\ophi^*\rangle$
into Eq.~\eqref{Ftermpotential} and setting $S=0$, the F-term potential reduces to
\begin{equation}\label{newFtermpotential}
V_F=\left|\kappa^2\left(M^2-H\oH\right)\right|^2
+2\,\frac{|\zeta|^2}{\Lambda^2}|\phi|^2|H|^2|\oH|^2
+\frac{|\zeta|^2}{\Lambda^2}|\phi|^4|H|^2
+\frac{|\zeta|^2}{\Lambda^2}|\phi|^4|\oH|^2\,.
\end{equation}
The upper panel of Fig.~\ref{hybrid} depicts the F-term scalar potential within the D-flat valley
for all model parameters set to unity.
Obviously, in the inflationary valley $S=H=\oH=0$ it has a flat inflaton direction $|\phi|$
and a tachyonic waterfall direction below some critical value $|\phi_{\text{c}}|$.

\subsection{Topological Defects}
One potential problem that arises if the waterfall is associated with the breaking of a non-Abelian unified gauge symmetry
$G$ is the possibility of
copiously producing topological defects~\cite{Kibble:1976sj} 
like magnetic monopoles in the waterfall transition at the end of inflation.
For such topological defects to form it is necessary that at the critical value when the waterfall occurs several different vacuum directions have degenerate masses and none is favored over the other. If the same vacuum is chosen everywhere in space, no topological defects can form.
In this respect, it is crucial to note that the VEV of the inflaton field already breaks the gauge symmetry $G$.
Due to this breaking, effective operators containing terms like $H^n \oH^m\phi^p \,\ophi^q$ can lead to a deformation of the potential
which can force the waterfall to happen in a particular field direction everywhere in space, avoiding the production of potentially problematic topological defects.
This is illustrated in the lower plot of Fig.~\ref{hybrid} for the Abelian example (even though no monopoles
can be created in this case; domain walls, however, can.).
We will discuss this in more detail in section~\ref{An Example: Sneutrino Inflation}.

\subsection{Radiative Corrections and Inflationary Predictions}

The tree-level flat direction is only lifted radiatively due to inflaton-dependent, SUSY breaking waterfall masses.
Diagonalizing the mass matrices in the $(H,\oH)$-basis, the eigenvalues calculated from
Eqs.~\eqref{toymodel} and~\eqref{newFtermpotential} are 1 Dirac fermion with
the squared mass $m_F^2=|\zeta|^2|\phi|^4/\Lambda^2$ and 2 complex scalars with squared
masses $m_S^2=|\zeta|^2|\phi|^4/\Lambda^2\pm|\kappa|^2M^2$.

Yet another potential problem may arise when the inflaton is a gauge non-singlet. It
is due to two-loop corrections to the inflaton potential which can induce a mass for the inflaton that is generically larger than the Hubble scale during inflation and would thus spoil slow-roll inflation~\cite{Dvali:1995fb}.
However, as we will discuss in section~\ref{Two-Loop Corrections}, due to the breaking of the
gauge symmetry during inflation these corrections to the inflaton potential
are not problematic in our model since they get suppressed by powers of the large gauge boson masses
induced by the inflaton VEV.

Since the two-loop corrections turn out to be negligible, it is enough to
consider the effective potential up to one-loop level when calculating 
predictions for the observable quantities.
In particular for a single field model as in the case $G=U(1)$, the relevant
inflationary predictions are the number of e-folds $N_{\text{e}}$ of inflation, the amplitude $P_{\mathcal{R}}$, spectral index $n_{\text{s}}$ and running spectral index $\dif n_{\text{s}}/\dif\ln k$ of the power spectrum for the scalar metric perturbations as well as
the tensor-to-scalar ratio $r$ giving the amplitude of the tensor metric perturbations.
These quantities can all be calculated from the potential
and its derivatives, more precisely from the slow-roll parameters given by~\cite{Liddle:2000cg}
\begin{equation}\label{slow-roll parameters}
\epsilon=\frac{1}{2}\left(\frac{V'}{V}\right)^2\,,\qquad
\eta=\left(\frac{V''}{V}\right)\,,\qquad
\xi^2=\left(\frac{V'V'''}{V^2}\right)\,.
\end{equation}
The number of e-folds from a given initial field value $\phi_{\text{i}}$ to the end of inflation can be calculated by
\begin{equation}
N_{\text{e}}=\int\limits_{\phi_{\text{e}}}^{\,\,\phi_{\text{i}}}\dif \phi \,\frac{1}{\sqrt{2\epsilon}}\,,
\end{equation}
where $\phi_{\text{e}}$ denotes the field value at the end of inflation. From this expression we can compute the field value $\phi_{60}$ 60 e-folds before the end of inflation, which is roughly the time when the relevant scales leave the horizon.
The other observables given in terms of the slow-roll parameters Eq.~\eqref{slow-roll parameters} read
\begin{equation}\label{observables}
n_{\text{s}}=1-6\,\epsilon+2\,\eta\,,\qquad
r=16\,\epsilon\,,\qquad
\frac{\dif n_{\text{s}}}{\dif \ln k}=16\,\epsilon\,\eta-24\,\epsilon^2-2\,\xi^2\,,
\end{equation}
while the amplitude of the scalar power spectrum has the form
\begin{equation}\label{amplitude power spectrum}
P^{1/2}_{\mathcal{R}}=\frac{1}{2\sqrt{3}\,\pi}\,\frac{V^{3/2}}{|V'|}\,,
\end{equation}
and all expressions have to be evaluated at $\phi = \phi_{60}$.
In section~\ref{One-Loop Corrections} we apply these formulae to give 
some typical predictions of our specific model treated in section~\ref{Sneutrino Inflation in Pati-Salam}.


\section{Sneutrino Inflation in SUSY Pati-Salam}\label{Sneutrino Inflation in Pati-Salam}

In this section we discuss a fully 
realistic example of SUSY hybrid inflation with a GNS inflaton where $G$ is identified with the SUSY Pati-Salam gauge group. Following the general ideas presented in the previous section, 
in the model under construction inflation will proceed along a trajectory in field space where the D-term contribution vanishes and the F-term contribution dominates the vacuum energy. In addition to that we want to associate the inflaton field to the ``matter sector'' of the theory so that the model is closely related to low energy particle physics. 
Typically if there are only matter fields in the (CP conjugated) right-handed Pati-Salam reps
$R_i^c$ this would lead to large D-term contributions incompatible with inflation. Therefore, in addition to the matter fields $R_i^c$ we also introduce another field $\oR^c$ in the conjugate rep of the gauge group. For simplicity, we will discuss here the case where $i=1,\dots,4$ and where there is only one $\oR^c$. As we will see, the introduction of $\oR^c$ is necessary in order to keep all the waterfall directions stabilized during inflation. The presence of $\oR^c$ also facilitates inflation to proceed along a D-flat valley. After inflation, one linear combination of the fields $R_i^c$ will pair with $\oR^c$ and become heavy, while three other combinations remain light and contain the three generations of SM fields. In addition, the superfields containing the right-handed neutrinos of the seesaw mechanism will obtain their large masses after inflation. In addition to the introduction of the model in this section, we also work out an example in full detail where the inflaton moves along a flat direction such that both $\oR^c$ and 
one of the $R_i^c$ get a VEV in the sneutrino direction.

\subsection{The Model}\label{The Model}
As an explicit realization of the idea of having a GNS inflaton, we consider the Pati-Salam gauge group $\PS$ \cite{Pati:1974yy}. For simplicity, we focus on the right sector of the theory only, i.e. fields that are charged under $SU(2)_{R}$.
From the point of view of the Higgs sector breaking PS to the SM this is sufficient, since
VEVs of one $(\rep{4},\rep{1},\rep{2})$ and one $(\bar{\rep{4}},\rep{1},\bar{\rep{2}})$
are enough for this purpose.
Therefore, let us first introduce the left-chiral right isospin doublet
leptoquark superfields and their conjugate rep given by
\begin{equation}\label{rightmatter}
\begin{split}
R^c_i&= (\bar{\rep{4}},\rep{1},\bar{\rep{2}})=
\begin{pmatrix}
u^c_i & u^c_i & u^c_i & \nu^c_i \\
d^c_i & d^c_i & d^c_i & e^c_i
\end{pmatrix}\,,\\
\oR^c &=(\rep{4},\rep{1},\rep{2})=
\begin{pmatrix}
\ou^c & \ou^c & \ou^c & \on^c \\
\od^c & \od^c & \od^c & \oel^c
\end{pmatrix}\,,
\end{split}
\end{equation}
where we have omitted color indices for convenience
and $i$ denotes a generation index. Here, the $R^c_i$ multiplets
contain the right-handed singlet fields under the SM gauge group.
The waterfall Higgs superfields breaking PS to the SM
after inflation reside in the multiplets
\begin{equation}\label{rightwater}
\begin{split}
H^c &= (\bar{\rep{4}},\rep{1},\bar{\rep{2}})=
\begin{pmatrix}
u_H^c & u_H^c & u_H^c & \nu_H^c \\
d_H^c & d_H^c & d_H^c & e_H^c
\end{pmatrix}\,,\\
\oH^c &=(\rep{4},\rep{1},\rep{2})=
\begin{pmatrix}
\ou_H^c & \ou_H^c & \ou_H^c & \on_H^c \\
\od_H^c & \od_H^c & \od_H^c & \oel_H^c
\end{pmatrix}\,.
\end{split}
\end{equation}
In addition, we introduce two further gauge singlet fields, namely $S$ and $X$.
The symmetry assignments to all the fields are given in the upper
half of Tab.~\ref{PSsymmetries}.
As we can see, we have introduced two additional symmetries:
a R-symmetry and a discrete $\Zten$ symmetry.
The lower half of Tab.~\ref{PSsymmetries} can be ignored
until we introduce the left doublets in a more general framework
in section~\ref{Left-Right Extension}.
We would also like to remark at this point that the symmetries and charge assignments of Tab.~\ref{PSsymmetries} are not unique and should mainly illustrate that it is possible to obtain the desired form of the superpotential by symmetry.

\begin{table}[!h]
\center
\begin{tabular}{c c c c}\hline
    & $\PS$                       & R & $\Zten$\\\hline
$S$ & $(\rep{1},\rep{1},\rep{1})$ & $1$ & $0$ \\
$X$ & $(\rep{1},\rep{1},\rep{1})$ & $0$ & $7$ \\
$H^c$ & $(\bar{\rep{4}},\rep{1},\bar{\rep{2}})$ & $0$ & $1$ \\
$\oH^c$&$(\rep{4},\rep{1},\rep{2})$ & $0$ & $2$\\
$R^c_{i}$&$(\bar{\rep{4}},\rep{1},\bar{\rep{2}})$& $1/2$ & $3$\\
$\oR^c$ & $(\rep{4},\rep{1},\rep{2})$ & $1/2$ & $4$\\ \hline
$H$ & $(\rep{4},\rep{2},\rep{1})$ & $0$ & $1$ \\
$\oH$ & $(\bar{\rep{4}},\bar{\rep{2}},\rep{1})$ & $0$ & $2$ \\
$L_i$ & $(\rep{4},\rep{2},\rep{1})$ & $1/2$ & $3$ \\
$\oL$ & $(\bar{\rep{4}},\bar{\rep{2}},\rep{1})$ & $1/2$ & $4$ \\ \hline
\end{tabular}
\caption{\label{PSsymmetries} Superfield content of the model and associated symmetries.}
\end{table}

Indeed, with the symmetry assignments of Tab.~\ref{PSsymmetries} the allowed terms in the superpotential up to dimension five operators are the following, 
\begin{equation}\label{super4c}
\begin{split}
W = &\,\kappa \,S\left( \frac{\langle X\rangle}{\Lambda}\,H^c \oH^c - M^2\right) \\
&+ \frac{\lambda_{ij}}{\Lambda}(R^c_i \oH^c) (R^c_j \, \oH^c)
+ \frac{\gamma}{\Lambda} (\oR^c H^c) ( \oR^c   H^c)
+ \frac{\zeta_{i}}{\Lambda}(R^c_i  \,\oR^c) (H^c  \oH^c)
+ \frac{\xi_{i}}{\Lambda}(R^c_i \oH^c)  (\oR^c H^c)   \, ,
\end{split}
\end{equation}
where two multiplets enclosed in brackets are contracted with their respective $SU(4)_C$ and $SU(2)_R$ indices. For simplicity, we only consider effective operators generated by the exchange of singlet messenger fields (for a detailed discussion see appendix~\ref{Effective Dimension 5 Operators in Pati-Salam}).

The roles of the superfields in this model are the following.
$S$ is the gauge singlet contributing the large
vacuum energy during inflation by its F-term, i.e $W_S \ne 0$.
It stays at zero both during and after inflation. A large mass for $S$ that keeps the field at zero can be generated by SUGRA effects due to higher order terms in the K\"ahler potential.
The right-doublets $H^c$, $\oH^c$ contain as scalar
components the waterfall fields which are zero
during inflation and become tachyonic
subsequently, ending inflation and breaking
$\PS$ to the MSSM by their VEVs.
The $SU(2)_{R}$-charged leptoquarks $R^c_{i}$ together
with $\oR^c$ provide the slow-roll inflaton directions
as scalar components.

After the end of inflation we want all components of three generations $R^c_i$, except for their right-handed neutrino to be light, whereas all components of $\oR^c$ need to be heavy. This is achieved by the introduction of several generations of $R^c_i$ fields.
With the number of generations of $R^c_i$ larger than the one of $\oR^c$ by three (e.g. $i=1,\ldots,4$),
all the $\oR^c$ fields pair up with some $R^c_i$ and form Dirac-type mass terms at the GUT scale and decouple from the theory.
Only three $R^c_i$ generations remain light.

Now we discuss the superpotential given in Eq.~\eqref{super4c} in more detail. The term proportional to $\zeta_i$ provides masses to all the components of the $H^c$ and $\oH^c$ fields during inflation when $R^c_i$ and $\oR^c$ get VEVs. Looking at the superpotential we can easily convince ourselves that without the presence of
the $\oR^c$ fields, not all of the waterfall squared masses are positive during inflation and their
immediate destabilization would not allow for slow-roll inflationary dynamics.
The introduction of the field $X$ is motivated as follows: We have imposed the discrete $\Zten$ symmetry to forbid a direct mass term for the $R^c_i$ and $\oR^c$ fields, therefore charging $R^c_i \,\oR^c$ under the symmetry. On the other hand, we have allowed the operator $R^c_i \,\oR^c \,H^c \oH^c $ in
Eq.~\eqref{super4c}, thus $H^c \oH^c$ cannot be invariant under this discrete symmetry.
Therefore, a superpotential term of the form $S\,H^c \oH^c$ is forbidden.
However, in the presence of the gauge singlet field $X$ that gets
a VEV around the Planck scale and breaks the discrete symmetry spontaneously, a similar term, $S\,\frac{X}{\Lambda}\,H^c \oH^c $, is allowed and it effectively generates the desired term after $X$ gets its VEV. 
To allow the term $S\,\frac{X}{\Lambda}\,H^c \oH^c $, the $X$ field carries a charge equal to the charge of the product $R^c_i \oR^c$ under the discrete symmetry, as can be seen in Tab.~\ref{PSsymmetries}.

\subsection{Inflationary Dynamics}
The inflationary epoch is determined by the scalar potential given by both F-term and D-term contributions of all chiral superfields. For the sake of simplicity, in this section we investigate only the global SUSY limit.

At the basic level, hybrid inflation requires a large vacuum energy density responsible for an exponential expansion of the scale factor and a nearly flat direction whose quantum fluctuations generate the metric perturbations.
In our model inflation proceeds along a trajectory in the field space of
$R^c_i$ and $\oR^c$ along which the D-term contributions vanish.
In such a D-flat valley, the F-term contribution from the $S$ field provides the necessary vacuum energy.
Both the $R^c_i$ and $\oR^c$ fields do not have any tree-level F-term mass contributions.
On the other hand, due to the large F-term contributions to the masses of the waterfall fields, they remain at zero during inflation (i.e. $H^c = \bar H^c = 0$). 
Therefore, in our PS framework
the tree-level F-term inflaton potential becomes
$V_F\sim \kappa^2 M^4$,
whereas the D-term potential reduces to
\begin{equation}\label{DtermpotPS}
V_D=\frac{g^2}{2}\,\sum_{a=1}^{18}\left({-}R^{c\dagger}_i \,\mathcal{T}^{a*}\,R^c_i+\oR^{c \dagger}\, \mathcal{T}^{a}\oR^c\right)^2\,.
\end{equation}
From now on, we denote the fifteen generators of $SU(4)_{C}$ by $\mathcal{T}^a$ ($a=1,\ldots,15$) which
have been explicitly listed in Tab.~\ref{PSgenerators} in appendix~\ref{Gauge Boson Masses}.
The three $SU(2)_{R}$ generators are as usual given in terms of the Pauli matrices
$\sigma^a/2$ which we refer to as $\mathcal{T}^{16},\ldots,\mathcal{T}^{18}$.
Furthermore, we assume $g\equiv g_{C}=g_{R}$ around the GUT scale.
Thus, the D-flatness conditions from Eq.~\eqref{DtermpotPS}
give the more specific conditions in the PS case
\begin{equation}\label{DflatconditionsPS}
R^{c\dagger}_{i}\mathcal{T}^{a*} R^c_{i}=\oR^{c \dagger} {\mathcal{T}^{a}} \oR^c\,,
\end{equation}
where the sum over all generations $i$ has to be taken into account
in each of the eighteen equations.
During inflation, our D-flat trajectory is thus constrained by the
conditions in Eq.~\eqref{DflatconditionsPS} which have to be imposed
on the F-term scalar potential.

Using Eq.~\eqref{DflatconditionsPS} it can be shown that several flat directions exist in this model. All these directions can in principle be valid trajectories for inflation to occur.
During inflation $R^c_i$ and $\oR^c$ acquire VEVs along one of these directions and break the PS symmetry.
The gauge fields coupled to this particular direction in field space become massive.
This direction is classically flat and lifted only by radiative corrections such that it is suitable for inflation.

On the other hand, other flat directions in field space along which the gauge symmetry is not broken and the gauge fields are still massless, acquire large two-loop mass contributions as will be clarified in section~\ref{Two-Loop Corrections}.
Such large mass contributions essentially lift these other flat directions strongly and drive their VEVs to zero.
After inflation the breaking is realized by the VEVs of $H^c$ and $\oH^c$.
In the next subsection we will explicitly consider inflation along the right-handed sneutrino directions $\nu^c$ and $\on^c$,
which provides one possible D-flat direction in field space.
We will show explicitly that in this case it happens generically that the VEVs of $H^c$ and $\oH^c$ are aligned in the right-handed sneutrino direction as well such that an example model of ``sneutrino inflation'' is realized with the inflaton being in a  non-singlet representation of $\PS$. It is important to emphasize that although the inflaton belongs to a non-singlet representation, it effectively behaves like a singlet since the $\PS$ gauge group is broken to the SM during inflation. As already mentioned, this will be important w.r.t.\ quantum corrections to the inflaton potential.

\subsection{An Example: Sneutrino Inflation} \label{An Example: Sneutrino Inflation}
As we have mentioned in the last section, the model has several tree-level flat directions in $R^c_i, \oR^c$ field space and in principle inflation can proceed along any of them. In this section we would like to discuss the inflationary dynamics when the inflaton fields acquire VEVs along the sneutrino direction. In addition we will also discuss the waterfall mechanism in more detail. It turns out to be an interesting feature of this particular flat direction that at the end of inflation, and for generic choices of parameters, the waterfall fields $H^c$ and $\oH^c$ acquire VEVs along the corresponding right-handed sneutrino directions $\nu^c_H$ and $\on^c_H$ as well.
We will also discuss how this preferred waterfall direction helps to avoid the production of topologically stable monopoles after inflation.

For an explicit example of the inflationary epoch we
consider a simple case where only one of the $R^c=R^c_1\ne 0$
is slow-rolling, while all the others remain at zero $R^c_{i\ne 1}=0$.
In addition, we want to realize inflation along the sneutrino direction, i.e.\
\begin{equation}\label{infvevs-1}
R^c=
\begin{pmatrix}
0 & 0 & 0 & \nu^c \\
0 & 0 & 0 &0
\end{pmatrix}\,,\qquad
\oR^c=
\begin{pmatrix}
0 & 0 & 0 & \on^c \\
0 & 0 & 0 & 0
\end{pmatrix}\,.
\end{equation}
This reduces our inflationary superpotential
in Eq.~\eqref{super4c} to the effective form
\begin{equation}\label{infsuperpotential-1}
\begin{split}
W_{\text{inf}}= &\kappa \,S\left(H^c\oH^c-M^2\right) \\
&+\lambda\, (\nu^{c}\, \on^{c}_H)^2
+\gamma\,(\on^{c}\,\nu^{c}_H)^2
+\xi\,(\nu^c\,\on^c)\,\nu^c_H \on^{c}_H
+\zeta\,(\nu^c\,\on^c)\,H^c \oH^c\,,
\end{split}
\end{equation}
where we have absorbed $\langle X\rangle / \Lambda$ into the definition of the other parameters.
Due to the VEVs in Eq.~\eqref{infvevs-1}, $\GPS$ is already broken to $\GSM$ during inflation.
If we can also ensure that the waterfall is forced into the $\nu^c_H$ and $\on^c_H$ directions in field space,
no monopoles will be produced after inflation.

Since $R^c$ and $\oR^c$ point in the right-handed sneutrino direction,
the D-term potential projects out only the part proportional to the
generator $\mathcal{T}^{15}$ of $SU(4)_{C}$ and $\mathcal{T}^{18}$ of $SU(2)_{R}$.
Hence, the global SUSY D-term potential reads
\begin{equation}
V_D=\frac{5}{16}\,g^2 \left(|\nu^c|^2-|\on^c|^2\right)^2\,.
\end{equation}
This potential obviously has a flat direction $|\nu^c|=|\on^c|$.
From now on, we assume that inflation occurs in this D-flat valley. Therefore the scalar potential during inflation has to be calculated  in the
inflationary trajectory $S=H^c=\oH^c=0$ with the D-flatness condition  $|\nu^c|=|\on^c|$ imposed.

For the D-flat direction $\langle\nu^c\rangle = \langle\on^c\rangle$ (assuming real VEVs), the field combination $\text{Re} (\delta\on^c - \delta \nu^c)$ having mass $5 \,g^2 \,{\langle\nu^c\rangle}^2/2$ is orthogonal to the flat direction $\text{Re} (\delta\on^c + \delta \nu^c)$ which remains massless.
On the other hand, for the other D-flat direction $\langle\nu^c\rangle = - \langle\on^c\rangle$, the field combination $\text{Re} (\delta \on^c + \delta \nu^c)$ acquires a mass of $5\, g^2 \,{\langle\nu^c\rangle}^2/2$ and is orthogonal to the flat direction $\text{Re} (\delta \on^c - \delta \nu^c)$.
The complete mass spectrum of the inflaton sector is listed in Tab.~\ref{gauge masses}.

Now we discuss how the waterfall mechanism works in our particular example.
We denote all complex scalar fields like their corresponding superfield
and decompose them into real and imaginary components as
$\nu^c_H=\left(\text{Re}(\tnu^c_H)+\imag\,\text{Im}(\tnu^c_H)\right)/\sqrt{2}$,
$\on^c_{H}=\left(\text{Re}(\tilde{\on}^c_{H})+\imag\,\text{Im}(\tilde{\on}^c_{H})\right)/\sqrt{2}$
and analogous for all the other waterfall fields.
Here and in the following, a tilde denotes canonically normalized fields and we define $\nu^c=|\tnu^c|/\sqrt{2}$ and
$\on^c=|\tnu^c|/\sqrt{2}$.

The full F-term potential contains the following terms
\begin{equation}\label{full F-term potential}
\begin{split}
V_F=&\left|\kappa\left(H^c\oH^c-M^2\right)\right|^2+\left|2\,\lambda\,(\nu^c)^2\on_H^c+\xi\,(\nu^c\,\on^c)\,\nu_H^c+\zeta\,(\nu^c\,\on^c)\,\nu_H^c\right|^2\\
&+\left|\kappa\,S\,\oH^c+\zeta\,(\nu^c\,\on^c)\,\oH^c\right|^2+\left|2\,\gamma\,(\on^c)^2\nu_H^c+\xi\,(\nu^c\,\on^c)\,\on_H^c+\zeta\,(\nu^c\,\on^c)\,\on_H^c\right|^2
\\
&+\left|\kappa\,S\,H^c+\zeta\,(\nu^c\,\on^c)\,H^c\right|^2+\left|2\,\gamma\,\on^c\,(\nu_H^c)^2+\xi\,\nu^c\,(\nu_H^c\on_H^c)+\zeta\,\nu^c\,(H^c\oH^c)\right|^2\\
&+\left|2\,\lambda\,\nu^c\,(\on_H^c)^2+\xi\,\on^c\,(\nu_H^c\on_H^c)+\zeta\,\on^c\,(H^c\oH^c)\right|^2\,,
\end{split}
\end{equation}
where terms containing single $H^c$ and $\oH^c$ superfields have to be summed over all
components. In terms like $(H^c\oH^c)$ all indices are contracted.

Due to large F-term contributions (cf. Eq.~\eqref{full F-term potential}) to their masses from the VEVs of the inflaton fields, the waterfall fields are fixed at zero during inflation.
However, as the inflaton fields slowly roll to smaller values, the masses of the waterfall fields decrease and finally one direction in field space  becomes tachyonic. The $H^c, \oH^c$ fields now quickly roll to their true minima and inflation ends by the ``waterfall''.
We now discuss in which direction in field space the waterfall will happen, i.e.\ which direction will become tachyonic first.

To start with, the masses of $(u^c_H, \ou^c_{H})$, $(d^c_H, \od^c_{H})$ and $(e^c_H, \oel^c_{H})$ obtain universal contributions from the terms with couplings $\kappa$ and $\zeta$.
For example, the normalized fields $\text{Re}\left(u^c_H+\ou^c_{H}\right)$
and $\text{Im}\left(\ou^c_{H}-u^c_H\right)$ acquire unstable squared masses $m_1^2=\frac{1}{4}\,|\zeta|^2\,|\tnu^c|^4-|\kappa|^2 M^2\,$, whereas the stable directions
$\text{Re}\left(\ou^c_{H}-u^c_H\right)$ and $\text{Im}\left(\ou^c_{H}+u^c_H\right)$ acquire squared masses $m_2^2=\frac{1}{4}\,|\zeta|^2\,|\tnu^c|^4+|\kappa|^2 M^2$.
Exactly the same mass spectra hold for $d^c_H$ and $e^c_H$.

However, due to the additional contributions from the non-universal couplings
$\lambda$, $\gamma$ and $\xi$, the SM-singlet directions $(\nu^c_H, \on^c_{H})$
obtain different masses.
Setting $\gamma=\lambda$,
we obtain the following eigenvalues for the scalars
\begin{equation}\label{nuscalareigen}
\begin{split}
m_{\text{Re}(\nu)1}^2&=\frac{|\zeta+\xi+2\gamma|
^2}{4}\,|\tnu^c|^4 -|\kappa|^2 M^2\,,\\
m_{\text{Re}(\nu)2}^2&=\frac{|\zeta+\xi-2\gamma|
^2}{4}\,|\tnu^c|^4 +|\kappa|^2 M^2\,.
\end{split}
\end{equation}
For the pseudoscalars, we obtain the eigenvalues
\begin{equation}\label{nupseudoeigen}
\begin{split}
m_{\text{Im}(\nu)1}^2&=\frac{|\zeta+\xi-2\gamma|
^2}{4}\,|\tnu^c|^4 -|\kappa|^2 M^2\,,\\
m_{\text{Im}(\nu)2}^2&=\frac{|\zeta+\xi+2\gamma|
^2}{4}\,|\tnu^c|^4 +|\kappa|^2 M^2\,.
\end{split}
\end{equation}
In Eqs.~\eqref{nuscalareigen} and \eqref{nupseudoeigen}, the first one can give rise to
an instability in both cases and corresponds to the directions
$\text{Re}(\on^c_{H}-\nu^c_H)$, $\text{Im}(\on^c_{H}-\nu^c_H)$, respectively.
The second, stable eigenvalues correspond to $\text{Re}(\on^c_{H}+\nu^c_H)$ and $\text{Im}(\on^c_{H}+\nu^c_H)$. All these masses are listed in Tab.~\ref{waterfall masses}.

One can easily calculate the critical values at which the system
gets destabilized by setting the dynamical masses to zero. We find
\begin{equation}
|\tnu^{c}_{\text{crit}}|=\sqrt{\frac{2\,|\kappa|\,M}{|\zeta|}}\,,
\end{equation}
for the $\text{Re}(u^c_H + \ou^c_{H})$, $\text{Im}(\ou^c_{H} - u^c_H),\,\ldots$ 
directions and the real, positive solutions
\begin{equation}
|\tnu_{\text{crit}}^c|=\sqrt{\frac{2\,|\kappa|\,M}{|\zeta+\xi+2\gamma|}}\,,\qquad
|\tnu_{\text{crit}}^c|=\sqrt{\frac{2\,|\kappa|\,M}{|\zeta+\xi-2\gamma|}}\,,
\end{equation}
for the $\text{Re}(\on^c_{H}-\nu^c_H)$- and $\text{Im}(\on^c_{H}-\nu^c_H)$-directions.

For generic non-zero values of $\gamma$ (and for example small $\xi$), either the $\text{Re}(\on^c_{H}-\nu^c_H)$- or the
$\text{Im}(\on^c_{H}-\nu^c_H)$-direction will become tachyonic for larger values of the inflaton VEV than
the $\text{Re}(u^c_H + \ou^c_{H}),\,\ldots$ directions. Consequently, it
destabilizes first and the waterfall will happen in this direction in field space.

We note that with the effective operators in Eq.~\eqref{super4c} included in this discussion, there is still the possibility of domain wall formation associated with the $\Ztwo$ symmetry $\nu^c_H \to - \nu^c_H$ and $\on^c_H \to - \on^c_H$. However additional effective operators at higher order that contain odd powers of $H^c$ and $\oH^c$ (in particular terms linear in $H^c$ and $\oH^c$) can efficiently lift this degeneracy and force the waterfall to occur in one unique direction.
An example for such a deformed inflationary potential is shown in section~\ref{Toy Model and Basic Ideas}.
For different possibilities to evade the cosmological domain wall problem, the
reader is referred to~\cite{Gelmini:1988sf}.  

In summary, since the gauge symmetry is already broken by the inflaton VEVs during inflation, higher-dimensional
operators allow to force the waterfall to happen in one single direction in field space such that a particular vacuum is chosen everywhere and the production of topological defects such as monopoles can be avoided.


\section{$\SOten$ SUSY GUTs}\label{Generalization to the $\SOten$ Framework}

We now turn to the embedding of the model into $\SOten$ GUTs. Starting with the model of the previous section, we will first make it explicitly left-right symmetric and then describe how its field content can be embedded in $\SOten$ representations. Consistency of the model with respect to one- and two-loop quantum corrections will be discussed in section~\ref{Radiative Corrections}.

\subsection{Left-Right Extension}\label{Left-Right Extension}
In order to make our simple example model of the previous section explicitly left-right-symmetric,
we need to add left-charged supermultiplets to the theory.
In addition to the right-charged matter fields and their conjugates,
defined in Eq.~\eqref{rightmatter}, we therefore introduce
left-chiral left-doublet leptoquarks contained in the multiplets
\begin{equation}
\begin{split}
L_i&=(\rep{4},\rep{2},\rep{1})=
\begin{pmatrix}
u_i & u_i & u_i & \nu_i \\
d_i & d_i & d_i & e_i
\end{pmatrix}\,,\\
\oL &=(\bar{\rep{4}},\bar{\rep{2}},\rep{1})=
\begin{pmatrix}
\ou & \ou & \ou & \on \\
\od & \od & \od & \oel
\end{pmatrix}\,,
\end{split}
\end{equation}
where we omitted the color indices for convenience
and $i$ denotes a generation index as before.
The waterfall Higgs-superfields breaking PS to the SM
by VEVs of their scalar components are given in Eq.~\eqref{rightwater}.
Making the field content left-right symmetric, we now have their
left-charged counterparts as well, which read
\begin{equation}
\begin{split}
H &=(\rep{4},\rep{2},\rep{1})=
\begin{pmatrix}
u_{H} & u_{H} & u_{H} & \nu_{H} \\
d_{H} & d_{H} & d_{H} & e_{H}
\end{pmatrix}\,,\\
\oH &=(\bar{\rep{4}},\bar{\rep{2}},\rep{1})=
\begin{pmatrix}
\ou_{H} & \ou_{H} & \ou_{H} & \on_{H} \\
\od_{H} & \od_{H} & \od_{H} & \oel_{H}
\end{pmatrix}\,.
\end{split}
\end{equation}

The symmetry assignments are given in Tab.~\ref{PSsymmetries}.
We note that at this stage the model contains two copies of the inflaton sector discussed in the previous section, one charged under $SU(2)_R$ and one charged under $SU(2)_L$, as well as additional couplings between the two sectors. In the absence of a discrete left-right symmetry we would expect the couplings in the left and right sector to be not exactly equal. With two potential sectors for inflation, inflation may happen in both of them with the respective sneutrinos playing the role of the inflaton. Thus we might have an ``inflaton race'' between the two sectors.
Once the waterfall happens in one of the two sectors (with different couplings in each sector we do not expect this to happen simultaneously), inflation ends since the vacuum energy given by the $F_S$-term vanishes.
At the same time the masses of the matter fields get fixed by the VEVs of the waterfall fields and the couplings between the left and the right sector.
When this happens, we \textit{(re)name} the sector in which the waterfall has occurred as the right sector under the SM gauge group. Before the breaking of PS to the SM the names right-charged and left-charged were arbitrary (referring with right-charged and left-charged to $SU(2)_R$ and $SU(2)_L$, respectively) and a renaming is always possible at this stage. Thus, without loss of generality, we can assume that PS is broken to the SM by the VEV of a right-charged PS Higgs field.


\subsection{Embedding into $\SOten$}
One attractive feature of $\SOten$ GUTs is that all matter fields of a family, including right-handed neutrinos, are contained in one $\sixteen$ representation of $\SOten$. If we furthermore consider a SUSY GUT, these fields are accompanied by their scalar superpartners. It is then tempting to try to realize inflation by one (or more) of the scalar fields belonging to such a
$\sixteen$ superfield.
In terms of the PS framework considered in the preceding sections, each of the left- and right-charged leptoquark superfields are unified into  $\sixteen$ reps and their conjugate counterparts into $\csixteen$ reps as
\begin{equation}\label{sixteendecomp}
\begin{split}
\sixteen &= (\rep{4},\rep{2},\rep{1})\oplus (\bar{\rep{4}},\rep{1},\bar{\rep{2}})\,,\\
\csixteen &= (\bar{\rep{4}},\bar{\rep{2}},\rep{1})\oplus (\rep{4},\rep{1},\rep{2})\,.
\end{split}
\end{equation}
In addition, the SM Higgs can be embedded into a $\rep{10}$ multiplet which
under PS decomposes as
\begin{equation}
\mathbf{10}= (\rep{1},\rep{2},\rep{2})\oplus(\rep{6},\rep{1},\rep{1})\,.
\end{equation}

In doing so, however, one immediately encounters a potential problem for realizing inflation, connected to
the Yukawa couplings of the matter representations to the $\rep{10}$ Higgs representation. If the theory contains renormalizable Yukawa interactions, i.e.\ terms of the form
\begin{equation} \label{SO10Yukawa}
y \, \sixteen . \rep{10} . \sixteen \, ,
\end{equation}
then the F-term of the $\rep{10}$ yields a contribution to the scalar potential
\begin{equation}
\sim \left| y \, \sixteen^2 \right|^2\,,
\end{equation}
that would represent quartic couplings of the inflaton field(s). Such a quartic term in the inflaton potential is, unless $y$ is extremely small, strongly disfavored by the WMAP CMBR data. On the other hand, in many flavor models based on GUTs combined with family symmetries, the Yukawa couplings, especially the ones for the first two families, do not arise from renormalizable couplings but rather from higher-dimensional operators. The suppression of the higher-dimensional operators allows to explain the hierarchical structure of the charged fermion masses. The Yukawa couplings are then generated after some family symmetry breaking Higgs field $\theta$, called flavon in the following, gets its VEV. Such Yukawa couplings can be schematically written as
\begin{equation}
y\, \frac{\langle \theta \rangle}{\Lambda} \, \sixteen . \rep{10}. \sixteen \, ,
\end{equation}
where ${\langle \theta \rangle}/{\Lambda}$ stands for the suppression of the Yukawa couplings by an effective operator and $\Lambda$ is the family symmetry breaking scale. It represents, in a simplified notation, the typically more complicated flavor sector of the theory, which is beyond the scope of the present paper.
As long as the flavon field $\theta$ obtains its VEV after inflation (and has zero VEV during inflation) the potentially problematic coupling in Eq.~\eqref{SO10Yukawa} is not appearing during inflation. We will assume this situation in the following.

The next issue we would like to address is how $\SOten$ gets broken down to the SM, and
how this breaking is connected to the monopole problem.
Since monopoles would be disastrous if they survived until today,
it is clear that either their production has to be avoided altogether (which is mandatory for phase transitions after inflation) or they have to be diluted by a subsequent stage of inflation.

The breaking of $\SOten$ can take place via various hierarchies of intermediate subgroups~\cite{Davis:1995bx}. One possibility, corresponding in some sense to the strategy followed so far in
this paper, is via the intermediate PS group
\begin{displaymath}
\SOten\,\longrightarrow \, \PS \, \longrightarrow \, \SM \,.
\end{displaymath}
In this breaking pattern, monopoles can in principle get produced in the first and in the second
stage of the breaking. In section~\ref{An Example: Sneutrino Inflation} we have already discussed how the monopole production
at the second stage of the breaking from PS to the SM can be avoided in our model of sneutrino inflation.
If we assume that $\SOten$ is broken to $\GPS$ before inflation, the monopoles produced at this stage of the breaking are diluted and thus unproblematic.

We would also like to note that the breaking via PS is not the only possible breaking pattern compatible
with GNS sneutrino inflation. For example, one could break $\SOten$ to the minimal left-right symmetric model
and then to the SM, avoiding monopole production completely at the second stage.
Since, apart from this, the discussion would be analogous to the breaking via PS, we will not dwell on this in any more detail.

Keeping these points in mind, let us now turn to the formulation of the model in the $\SOten$ framework. As described above, we unify the left- and right-charged multiplets into $\sixteen$'s and $\csixteen$'s (cf.\ Eq.~\eqref{sixteendecomp}).
The matter fields containing the SM fermions and their superpartners will be denoted as $F_i=\sixteen_i$ and $\oF=\csixteen$. The ``waterfall'' Higgs fields are unified into the $\SOten$ representations $H=\sixteen$ and $\oH=\csixteen$.
The symmetry assignments are basically chosen as in the previous sections. An example superfield content with associated symmetry assignments is displayed in Tab.~\ref{SO(10)symmetries}.
\begin{table}[!h]
\center
\begin{tabular}{c c c c c}
\hline
     & $SO(10)$ & R & $\Zten$ & $\Ztwo$\\ \hline
$S$ & $\mathbf{1}$ & $1$ & $0$ & $+$\\
$X$ & $\mathbf{1}$ & $0$ & $7$ & $+$\\
$H$ & $\sixteen$ & $0$ & $1$ & $+$\\
$\oH$&$\csixteen$ & $0$ & $2$ & $+$\\
$F_{i}$&$\sixteen$& $1/2$ & $3$ & $+$\\
$\oF$ & $\csixteen$ & $1/2$ & $4$ & $+$\\ \hline
$h$ & $\mathbf{10}$ & $0$ & $4$ & $-$\\
$\theta$ & $\mathbf{1}$ & $0$ & $0$ & $-$\\ \hline
\end{tabular}
\caption{\label{SO(10)symmetries} Example of $SO(10)$ superfield content and associated symmetries.}
\end{table}
Up to dimension seven operators, the allowed terms in the superpotential read
\begin{equation}
\begin{split}
W=&\kappa \, S\left(\frac{\langle X\rangle}{\Lambda}\,H\oH-M^2\right)\,+\,\frac{\lambda_{ij}}{\Lambda}\,F_iF_j\oH\oH
\,+\,\frac{\gamma}{\Lambda}\,\oF\oF H H \,+\,\frac{\zeta_i}{\Lambda}\,F_i\oF H\oH \\
&\,+\, y_{ij}\,\frac{\langle\theta\rangle}{\Lambda}\,F_i \,h\, F_j
\,+\, \tilde{y}\,\frac{\langle\theta\rangle}{\Lambda^3}\, h^2\oF\,h\,\oF \,+\,\ldots\,,
\end{split}
\end{equation}
where $h=\rep{10}$ contains the SM Higgs superfields. Like in the PS version of the model, we assume that $X$ has already acquired its large VEV $\langle X\rangle \sim \Lambda$ before inflation has started. Furthermore we assume $\langle \theta\rangle =0$ during inflation as explained above.

The part of the superpotential of our model relevant for inflation has the form
\begin{equation}
W_{\text{inf}}=\kappa \, S\left(H\oH-M^2\right)\,+\,\frac{\lambda_{ij}}{\Lambda}\,F_iF_j\oH\oH
\,+\,\frac{\gamma}{\Lambda}\,\oF\oF H H \,+\,\frac{\zeta_i}{\Lambda}\,F_i\oF H\oH
\,+\,\ldots\, \; .
\end{equation}
We assume that $SO(10)$ is broken to $\GPS$ before inflation and then inflation, as well as the waterfall after inflation
are realized as discussed in section~\ref{Sneutrino Inflation in Pati-Salam}.

We would like to emphasize at this point that the minimalist field content and the choice of symmetries mainly serves the purpose of giving a proof of existence that GNS inflation can be realized in $\SOten$ GUTs.
In a fully realistic model, which e.g.~may also contain a full flavor sector, different symmetries may have to be chosen and the field content may have to be extended.


\section{Radiative Corrections}\label{Radiative Corrections}
In this section, we describe the radiative corrections
to the flat tree-level inflaton potential.
Subsection~\ref{One-Loop Corrections} is dedicated to
the one-loop Coleman-Weinberg corrections \cite{Coleman:1973jx}.
We summarize the full mass spectrum during inflation as
calculated in detail in appendix~\ref{Mass Spectrum during Inflation}
and section~\ref{An Example: Sneutrino Inflation}.
As it turns out, in the absence of SUGRA masses for the gauginos,
only the fields of the waterfall sector show a splitting between the masses
of the scalar and fermionic components
and hence contribute to the lifting of the
flat direction at one-loop level.
In subsection~\ref{Two-Loop Corrections}, we give estimates
for the potentially dangerous two-loop corrections pointed out in~\cite{Dvali:1995fb} and
show that they are small and can be neglected in our model.

\subsection{One-Loop Corrections}\label{One-Loop Corrections}
Typically, the tree-level flat directions get lifted by the Coleman-Weinberg
one-loop radiative corrections to the effective potential given by
\begin{equation}\label{onelooppotential}
V_{\text{loop}}(R^c_i)=\,\frac{1}{64\,\pi^2}\,\text{Str} \left[\mathcal{M}^4(R^c_i)
\left(\ln{\left(\frac{\mathcal{M}^2(R^c_i)}{Q^2}\right)}
-\frac{3}{2}\right)\right]\,,
\end{equation}
where $Q$ is a renormalization scale.
Since the supertrace is taken over all fermionic
and bosonic DOFs, we have to calculate the full mass spectrum.

In our previous studies we have already calculated the one-loop
contributions to the inflaton potential due to the inflaton field-dependent masses of the scalar and fermionic
components of the waterfall sector superfields. The calculation here can
be performed analogously.
In addition to the waterfall sector we have to consider the gauge sector of
the theory for the one-loop contributions, i.e.\ the loop contributions from
inflaton field-dependent masses of gauge bosons and gauginos.
Plugged into Eq.~\eqref{onelooppotential}, we end up with the effective potential
in our model.

Let us start with the gauge sector masses of our model, since
we will see that under our assumptions, SUSY-breaking does not directly affect this sector.
The aforementioned assumptions contain the absence of a direct SUGRA gaugino mass term
\begin{equation}
\mathcal{L}_{\text{gaugino}}= \frac{1}{4}\,\MP\,\text{e}^{-\langle G\rangle/(2\MP^2)}
\left\langle G^l\left(G^{-1}\right)^k_l\,\frac{\partial f^*_{ab}}{\partial \phi^k}\right\rangle
\,\lambda^a\,\lambda^b\,+\,\text{h.c.}\,,
\end{equation}
where $G$ denotes the K\"ahler function defined as $G=K+\ln|W|^2$.
The presence (or absence) of this contribution to gaugino masses depends on the details of the SUGRA model.
If, for instance,  the gauge kinetic function is diagonal and constant $f_{ab}=\delta_{ab}$ (or, more precisely, independent of fields that obtain a non-zero F-term such as $S$ in our model) then the contribution vanishes.
In the following we will assume this situation for simplicity.
Tab.~\ref{gauge masses} summarizes the mass eigenvalues
of the gauge bosons, the gaugino-chiral fermion mixings and the
D-term real scalars. Obviously, the supertrace over these contributions
vanishes and they do not contribute to Eq.~\eqref{onelooppotential}.

\begin{table}[!h]
\center
\begin{tabular}{l c c}\hline
Quantum Fields  & \qquad & Squared Masses $m^2$\\ \hline
8 gauge bosons & \qquad & $g^2\,{\langle\nu^c\rangle}^2$ \\
1 gauge boson & \qquad & $5\,g^2\,{\langle\nu^c\rangle}^2/2$\\
\hline
8 Dirac fermions & \qquad & $g^2\,{\langle\nu^c\rangle}^2$ \\
1 Dirac fermion & \qquad & $5\,g^2\,{\langle\nu^c\rangle}^2/2$ \\
\hline
8 real scalars & \qquad & $g^2\,{\langle\nu^c\rangle}^2$ \\
1 real scalar & \qquad & $5\,g^2\,{\langle\nu^c\rangle}^2/2$ \\
\hline
\end{tabular}
\caption{\label{gauge masses} Gauge sector mass spectrum.}
\end{table}

Hence, the $\langle\nu^c\rangle$-dependent, SUSY-breaking contributions
arise from the waterfall sector masses only.
Their squared masses are displayed in Tab.~\ref{waterfall masses}.
These masses carry the SUSY-mass splittings $\mu=\kappa\,M$
and thus contribute to the one-loop inflaton potential via Eq.~\eqref{onelooppotential}, lifting the
tree-level flat direction.
\begin{table}[!h]
\center
\begin{tabular}{l c c}\hline
Quantum Fields  & \qquad & Squared Masses $m^2$\\ \hline
7 Dirac fermions & \qquad & $|\zeta|^2\,{\langle\nu^c\rangle}^4$ \\
1 Majorana fermion & \qquad & $|2 \,\gamma - \zeta - \xi|^2\,{\langle \nu^c \rangle}^4$ \\
1 Majorana fermion & \qquad & $|2 \,\gamma + \zeta + \xi|^2\,{\langle \nu^c \rangle}^4$ \\
\hline
7 complex scalars & \qquad & $|\zeta|^2\,\langle\nu^c\rangle^4-|\kappa|^2 M^2$ \\
7 complex scalars & \qquad & $|\zeta|^2\,\langle\nu^c\rangle^4+|\kappa|^2 M^2$ \\
1 real scalar & \qquad & $|\zeta+\xi-2\gamma|^2\,\langle\nu^c\rangle^4 +|\kappa|^2 M^2$ \\
1 real scalar & \qquad & $|\zeta+\xi-2\gamma|^2\,\langle\nu^c\rangle^4 - |\kappa|^2 M^2$\\
1 real scalar & \qquad & $|\zeta+\xi+2\gamma|^2\,\langle\nu^c\rangle^4 + |\kappa|^2 M^2$ \\
1 real scalar & \qquad & $|\zeta+\xi+2\gamma|^2\,\langle\nu^c\rangle^4 -|\kappa|^2 M^2$ \\
\hline
\end{tabular}
\caption{\label{waterfall masses} Waterfall sector mass spectrum.}
\end{table}

\begin{figure}[!h]
\psfrag{x}{\hspace{-0.6cm}$\langle\tilde{\nu}^c\rangle\, [\MP]$}
\psfrag{V}{\hspace{-1cm}$V_{\text{eff}}\,[10^{-13} \MP^4]$}
\center
\includegraphics[scale=0.9]{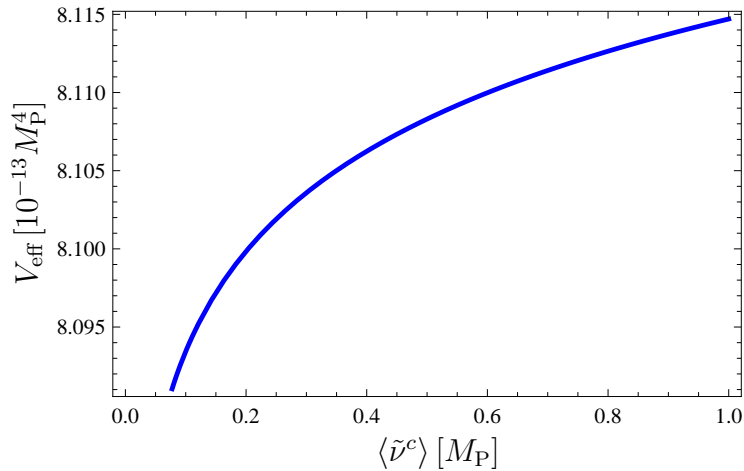}
\caption{\label{loopplot}Coleman-Weinberg corrected inflaton potential. The negative curvature of the potential gives rise to a red tilted spectral index as observed by WMAP~\cite{Komatsu:2010fb}.}
\end{figure}

For an example set of parameters $\kappa=\xi=0.1$, $\gamma=-0.1$, $\zeta=0.2$
and mass scale $M=0.003\,\MP$ as well as a renormalization scale
$Q=\sqrt{2}\,M$, we have plotted the one-loop effective potential in Fig.~\ref{loopplot}.
It has the typical shape of the Coleman-Weinberg potential in hybrid
inflation. Since in the case considered here the inflationary trajectory is a straight line in field space we are effectively dealing with a single-field model and the inflationary predictions can be directly calculated using equations~\eqref{slow-roll parameters}-\eqref{amplitude power spectrum}. 
The negative curvature of the potential gives rise to a spectral index below one (typically $n_{\text{s}}\sim 0.98$), as well as a small tensor-to-scalar ratio $r\lesssim10^{-2}$. The COBE normalization $P^{1/2}_{\mathcal{R}}\sim 5\cdot 10^{-5}$ fixes the scale $M$ and we have assumed $N_{\text{e}}=60$. Furthermore, we do not expect large non-gaussianities since as mentioned above the inflationary trajectory is not curved in field space
\footnote{We note that for more complicated trajectories, non-gaussianities may arise.}.

We note that the prediction for $n_{\text{s}}$ can be further lowered and thus brought even closer to the best fit value of the latest WMAP results~\cite{Komatsu:2010fb}, when the possible K\"ahler potential coupling between the $S$ field and the waterfall fields is taken into account~\cite{Antusch:2009ef}\footnote{We note that a lower spectral index in SUSY hybrid inflation models can also be achieved by different means~\cite{Rehman:2009nq}.}.

\subsection{Two-Loop Corrections}\label{Two-Loop Corrections}
In this section, we discuss how the two-loop
Dvali problem~\cite{Dvali:1995fb} is not endangering
inflation in our type of models. First of all, we will state the problem in general terms and later
show how such two-loop corrections are suppressed in our case.

For a GNS inflaton there are two basic conditions that have to be fulfilled to
give rise to the problem.
First of all, there is one superfield $S$,
which contributes the large vacuum
energy by its F-term $W_S\ne 0$.
Secondly, this superfield has to be coupled to some
non-singlet superfields, in our case $H^c$, $\oH^c$.
A relevant superpotential term reads, for example,
\begin{equation}
W\supset \kappa \,S\left(H^c\oH^c-M^2\right)\,.
\end{equation}

\begin{figure}[!h]
\psfrag{H}{$H^c$, $\oH^c$}
\psfrag{I}{$\phi$}
\psfrag{A}{$A_{\mu}$}
\psfrag{B}{$A^{\mu}$}
\center
\includegraphics{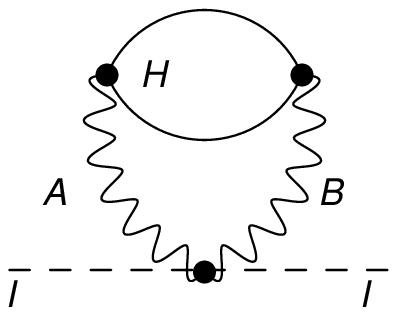}
\hspace{0.4cm}
\includegraphics{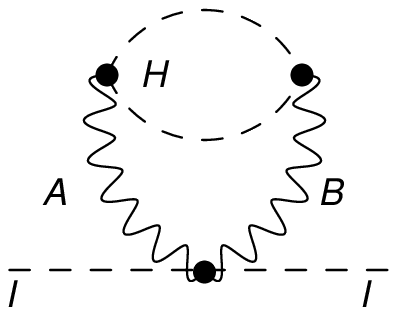}
\hspace{0.4cm}
\includegraphics{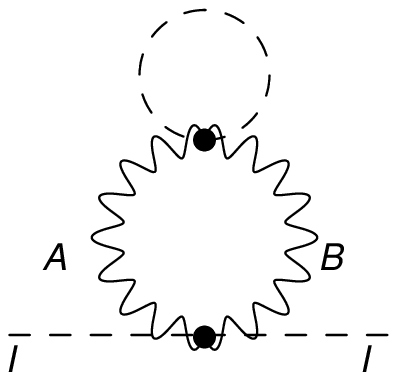}\\[0.7cm]
\psfrag{P}{$\delta\nu^+$}
\psfrag{M}{$\delta\nu^-$}
\psfrag{p}{$\delta\nu_H^+$}
\psfrag{m}{$\delta\nu_H^-$}
\center
\includegraphics[scale=0.8]{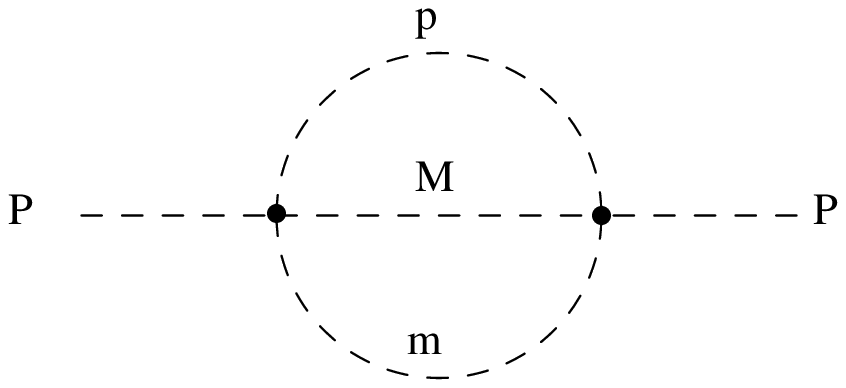}
\hspace{0.4cm}
\psfrag{X}{$\psi_{H^c}$}
\psfrag{n}{$\phi$}
\psfrag{L}{$\lambda^a$}
\psfrag{h}{$H^c,\oH^c$}
\psfrag{H}{$\psi_{\phi}$}
\includegraphics[scale=1]{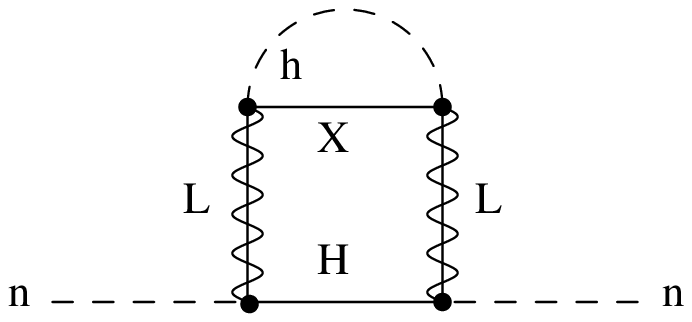}\\[0.7cm]

\caption{\label{twoloopdiag} Two-loop diagrams contributing to the gauge $\eta$-problem (Dvali problem) pointed out in~\cite{Dvali:1995fb}.
In the fourth diagram, we have defined mass eigenstates
$\delta\nu^+=\text{Re}(\delta\on^c+\delta\nu^c)$,\,
$\delta\nu^-=\text{Re}(\delta\on^c-\delta\nu^c)$,\,
$\delta\nu^+_H=\text{Re}(\delta\on^c_H+\delta\nu^c_H)$ and
$\delta\nu^-_H=\text{Re}(\delta\on^c_H-\delta\nu^c_H)$.
}
\end{figure}

If these premises are given, any gauge non-singlet
direction $\phi$ will receive two-loop contributions to its effective
mass of the order
\begin{equation}\label{twoloopmass}
{\delta m}^2\sim \,\frac{g^4}{(4\pi)^4}\,\frac{|W_S|^2}{m_{F}^2}\,,
\end{equation}
where $g$ is the gauge coupling constant and $m_{F}$ refers to
the SUSY conserving mass of the $H^c$, $\oH^c$ superfields.
In Fig.~\ref{twoloopdiag}, we have displayed the diagrams contributing to the mass correction.

Typically, a contribution as in Eq.~\eqref{twoloopmass} is large enough to provide
an inflaton mass such that $\delta m>\mathcal{H}$, 
which denotes the Hubble scale during inflation, and thus
slow-roll inflation is spoiled. Hence, the Dvali problem can in some sense
be considered a ``two-loop gauge $\eta$-problem'' since it implies
$|\eta|\sim 1$ due to radiative corrections from gauge interactions.
In our simple model given in section~\ref{The Model},
we are thus interested in $\phi=\{R^c, \oR^c\}$.

However, Eq.~\eqref{twoloopmass} cannot be applied to our model since
the inflaton VEV already breaks the gauge symmetry $\GPS$ during inflation.
Indeed, Eq.~\eqref{twoloopmass} is calculated under the assumption that the gauge bosons $A_{\mu}$
mediating the loops are massless, which is not the case in our model.
As we will now argue, the broken gauge symmetry during inflation corresponds to large gauge boson masses
that suppress the two-loop contributions of Fig.~\ref{twoloopdiag}.

More explicitly, for $\phi=\{\nu^c,\on^c\}$, the gauge bosons in Fig.~\ref{twoloopdiag}
are contained in the coset $\GPS/\GSM$ corresponding to the massive ones,
which is why their contributions get suppressed.
Another way to say this is that the effective gauge symmetry
during inflation is $\GSM$ under which the inflaton direction $\phi$ is a singlet.
All the other directions $\phi=\{u^c, d^c, e^c, \ou^c, \od^c, \oel^c\}$ couple to
gauge bosons that are still massless, which allows the use of Eq.~\eqref{twoloopmass}.
As a consequence, they just obtain additional mass contributions helping
to keep them at zero during the inflationary epoch.

Let us now estimate the typical size of the two-loop corrections
in our model in the large gauge boson mass limit $M_g\gg p$.
For the SUSY-splitted waterfall masses, we have plugged in
\begin{equation}
m^2_+=m^2_F+\mu^2\,,\qquad
m^2_-=m^2_F-\mu^2\,,
\end{equation}
where $m^2_F\sim\zeta^2{\langle\nu^c\rangle}^4/\MP^2$ is the
mass of the waterfall superpartner chiral fermion and $\mu=\kappa\, M$
is the SUSY-breaking scale.
Due to the non-renormalization theorem,
all contributions not proportional to powers of $\mu$ must cancel such that
in the SUSY-limit $\mu\rightarrow 0$ the total two-loop contribution vanishes.
Thus, we expand the final loop-integrals in terms of $\mu$.

In analogy to the calculations in~\cite{AlvarezGaume:1981wy}
we find that in the large gauge boson mass limit we obtain that the diagrams in Fig.~\ref{twoloopdiag} lead to 
two-loop mass contributions of the order
\begin{eqnarray}\label{twoloopmass1}
{\delta m}^2&\sim&\frac{g^4}{(4\pi)^4}\,\frac{m^2_F\,\mu^4}{M^4_g}\,, \\
{\delta m}^2&\sim& \frac{g^4}{(4\pi)^4}\,\frac{\mu^4}{M_g^2}\,,\\
{\delta m}^2&\sim& \frac{g^4}{(4\pi)^4}\,\frac{m_F\,\mu^4}{M_g^3}\,.
\end{eqnarray}
Using the values $\kappa=0.05$, $\zeta=0.2$, $g=0.5$, $M=3.4\cdot 10^{-3}\MP$ and
$\langle\nu^c\rangle = 0.36\,\MP$ at about $50$ e-folds before the end of inflation,
taken from Ref.~\cite{Antusch:2008pn} where a similar effective
superpotential has been analyzed,
we can further estimate
\begin{eqnarray}
\frac{{\delta m}^2}{\mathcal{H}^2}&\sim& \frac{3 \,\zeta^2\kappa^2}{(4\pi)^4}\sim\mathcal{O}(10^{-8})\,,\\
\frac{{\delta m}^2}{\mathcal{H}^2}&\sim& \frac{3\,g^2\kappa^2}{(4\pi)^4}\left(\frac{\MP}{\langle\nu^c\rangle}\right)^2
\sim\mathcal{O}(10^{-6})\,,\\
\frac{{\delta m}^2}{\mathcal{H}^2}&\sim& \frac{3\,g\,\zeta\,\kappa^2}{(4\pi)^4}\left(\frac{\MP}{\langle\nu^c\rangle}\right)
\sim\mathcal{O}(10^{-7})\,.
\end{eqnarray}
The Hubble scale during inflation is given by $\mathcal{H}^2\sim \kappa^2 M^4/3\MP^2$.
We can thus conclude that the two-loop contributions can be neglected in our model.


\section{Generalization to Supergravity}\label{Generalization to Supergravity}

So far, we have investigated the proposed model within
a global SUSY framework only.
The purpose of this section is to outline how GNS inflation 
can be generalized to local SUSY (i.e.\ SUGRA).
When dealing with inflation model building in SUGRA, a typical
problem that arises and with which one has to cope is the
$\eta$\,-\,problem. In section~\ref{eta-Problem}, we shortly
review how this generally threatens the flatness of inflaton potentials
in SUGRA. A possible solution to the $\eta$\,-\,problem in
SUGRA is the use of a fundamental symmetry in the K\"ahler potential, 
for example of a Heisenberg symmetry. 
A brief summary of the Heisenberg symmetry approach and how one can
apply it to our type of model follows in section~\ref{Heisenberg Symmetry Solution}.

\subsection{The $\eta$\,-\,Problem}\label{eta-Problem}
From the effective field theory point of view, for any (singlet or gauge non-singlet) field $F$,  
operators like
\begin{equation}\label{dangerousoperators}
V_0\,\frac{\left({F}^\dagger F\right)^n}{\MP^{2n}}\,,
\end{equation}
can be written in the potential, where $V_0$ is the vacuum energy.
However, the first term in such an expansion ($n=1$)
induces a large contribution to the inflaton mass proportional
to $V_0$, i.e.\ $V'' \sim V_0/\MP^2$.
Plugged in the formula for the slow-roll parameter $\eta=\MP^2\left(V''/V\right)$,
this generically spoils inflation due to a leading contribution $\eta\sim 1$.

Within SUGRA theories, this so-called $\eta$\,-\,problem
typically appears, since gravity couples to everything and thus
also induces a coupling of all the fields to the vacuum energy density $V_0$.
Especially in the F-term contribution to the scalar potential given by
\begin{equation}\label{SUGRAFtermpotential}
V_F=\text{e}^{K}\left(K^{ij*}D_i W\,D_{j*}W^*-3|W|^2\right)\,,
\end{equation}
this is obvious, since for a minimal K\"ahler potential
$K=F^\dagger F$ giving rise to canonical kinetic terms,
an expansion of the exponential in Eq.~\eqref{SUGRAFtermpotential}
leads to a scalar potential of the form
\begin{equation}\label{SUGRAFtermexpansion}
V_F\sim\left(1+\frac{F^\dagger F}{\MP^2}+\ldots\right)V_0\,.
\end{equation}
When we compare Eq.~\eqref{SUGRAFtermexpansion} to Eq.~\eqref{dangerousoperators},
it is exactly these dangerous terms that reappear in the F-term potential of a SUGRA theory.
This states the $\eta$\,-\,problem of SUGRA inflation~\cite{Copeland:1994vg}.

\subsection{Heisenberg Symmetry Solution}\label{Heisenberg Symmetry Solution}
In order to embed our model into a SUGRA
framework, we have to solve the $\eta$\,-\,problem.
Therefore, in addition to the superpotentials treated so far,
we may introduce a K\"ahler potential, as proposed in~\cite{Antusch:2008pn},
that is invariant under a Heisenberg symmetry~\cite{Binetruy: 1987}. 
In this approach an additional (``modulus'') field $T$ is introduced. 
The Heisenberg symmetry~\cite{Binetruy: 1987} given by
the non-compact Heisenberg group transformations
\begin{equation}\label{Heisenberg transformations}
T\rightarrow T+\text{i}\,\beta\,,\qquad
T\rightarrow T+\alpha^{* I}\,F_I+\frac{|\alpha_I|^2}{2}\,,\qquad
F_I\rightarrow F_I+\alpha_I\,,
\end{equation}
gives rise to the invariant combination
\begin{equation}\label{Heisenberg}
\rho=T+T^*-F_{i}^{\dagger}F_{i}-{\oF}^{\dagger}\oF\,,
\end{equation}
where the $\alpha_I$ and $\beta$ are infinitesimal transformation parameters.
Note that the index $I$ runs over all generation indices, gauge indices and
representations (i.e.~also $\oF$).

Following \cite{Antusch:2008pn}, a suitable Heisenberg symmetry invariant K\"ahler potential is given by 
\begin{equation}\label{Kaehlerpotential}
K\,=\,k(\rho) \,+\, \left(\,1\,+\,\kappa_{S}\,|S|^2\,+\,\kappa_{\rho}\,\rho\,\right)|S|^2
\,+\,H^{\dagger}H\,+\,\oH^{\dagger}\oH \,+\, h^{\dagger}h\,,
\end{equation}
where the dagger indicates complex conjugation and summation over
all gauge indices.
Note that the function $k(\rho)$ can be a general function which is only
constrained by the requirement that the resulting potential has
a stable minimum $\rho_{\text{min}}$ in which $\rho$ can settle and that $k'(\rho_{\text{min}})<0$
to obtain positive kinetic terms for the inflaton fields.
An important feature of Eq.~\eqref{Kaehlerpotential} is the term $\kappa_S |S|^4$.
For negative $\kappa_S$, this gives a large mass to the $S$ field which
stabilizes it at zero during inflation
(which has been assumed throughout the paper so far).

We would like to note at this point that the Heisenberg symmetry is not meant to be an
exact symmetry of the theory, but rather an approximate one. It is even
necessary to break the Heisenberg symmetry at some level since otherwise the inflaton potential
would be exactly flat and inflation could not end. In our model, the Heisenberg symmetry is broken 
by effective operators in the superpotential (which conserve tree-level flatness but induce a slope of
the inflaton potential at loop level) as well as by the gauge interactions. At tree-level, the latter effects vanish in the D-flat valley and the gauge loop effects have been discussed in detail in section \ref{Radiative Corrections}. Thus, the breaking of 
the Heisenberg symmetry in our scenario is capable of generating the desired slope of the inflaton potential but does not
endanger the solution to the $\eta$-problem.

If we choose $\rho$ and the components of $F_{i}$
and $\oF$ to be the independent degrees of freedom (DOFs)
and eliminate the $T$-DOFs, the F-term potential
in the inflationary minimum $S=H=\oH=h=0$
is of the form
\begin{equation}\label{modulus potential}
V_F\sim \kappa^2 M^4\,\frac{\euler^{k(\rho)}}{(1+\kappa_\rho\,\rho)}\,,
\end{equation}
and thus flat at tree-level in direction of the $F_{i}$ and $\oF$
components.
As can be seen from Eq.~\eqref{modulus potential},
the additional coupling $\kappa_\rho$ in the K\"ahler potential
is essential to stabilize the modulus field $\rho$.
This is possible for negative $\kappa_\rho$.

In a SUGRA framework under the assumption of a constant diagonal
gauge kinetic function $f_{ab}=\delta_{ab}$, the D-term potential
will also be $\rho$-dependent and of the form
\begin{equation}
V_D\sim\frac{g^2}{2}\,k'(\rho)^2\sum_{a}\left(F^{\dagger}_i \,\mathcal{T}^{a}F_i-\oF^{\dagger}\, \mathcal{T}^{a*}\oF\right)^2\,.
\end{equation}
The basic difference to the global SUSY D-term contribution in Eq.~\eqref{DtermpotPS}
is the global factor of $k'(\rho)^2$.
Due to the fact that the modulus quickly acquires its minimum at the
very beginning of inflation from Eq.~\eqref{modulus potential}, 
$k'(\rho_{\text{min}})^2$ soon approaches a constant value and
the D-flatness conditions basically do not change w.r.t. the
global SUSY ones.

At this point we would like to emphasize the special properties of the superpotential of our model, i.e Eq. \eqref{super4c}. In our setup the inflationary superpotential vanishes during inflation and the vacuum energy originates from the F-term of some field other than the inflaton. It has recently been pointed out in \cite{Antusch:2009vg} that, due to this property, the class of models considered here for GNS inflation is generically very suitable for the generalization from global SUSY to SUGRA. 

We furthermore emphasize that the Heisenberg symmetry approach 
is especially suitable for solving the $\eta$-problem for GNS inflation in SUGRA, in contrast to other approaches applicable to gauge-singlet inflation. For example, in~\cite{Antusch:2009ef,Mooij:2010cs} a shift symmetry in the K\"ahler potential has been used to solve the $\eta$-problem in a similar class of inflation models but with a gauge-singlet inflaton field. Clearly, a shift symmetry $\phi\rightarrow \phi+\text{i}\,\mu$ cannot
be applied to GNS inflation since it does not respect gauge symmetry.

In summary, the use of a Heisenberg symmetry in the K\"ahler potential
is particularly suitable for realizing GNS inflation in
SUGRA, because it allows to solve the SUGRA $\eta$-problem in a way that is compatible with a charged inflaton.


\section{Summary and Conclusions}
In this paper we have explored the novel possibility that, within SUSY GUTs, the inflaton responsible for
cosmological inflation is a gauge {\it non}-singlet under some gauge group $G$.
For definiteness we have considered SUSY hybrid inflation
where we have shown that the scalar components of gauge non-singlet superfields,
together with fields in conjugate representations, may form a D-flat direction suitable for inflation.

We have first sketched an explicit example of this scenario based on the Abelian gauge group $G=U(1)$.
We then presented a realistic model of this kind based on the relevant part of the SUSY Pati-Salam gauge group under which the inflaton transforms, namely $G=SU(4)_C\times SU(2)_R$. In such a framework we have shown how it is possible for the inflaton to consist of the scalar components of a pair of gauge non-singlet matter superfields $R^c$ and $\oR^c$, in conjugate representations under the Pati-Salam gauge group, rolling along a D-flat valley and coupled to a pair of gauge non-singlet Higgs superfields $H^c$ and $\oH^c$, also in conjugate representations of the same gauge group, whose VEVs end inflation, according to the hybrid inflation scenario, breaking the Pati-Salam gauge group. We emphasize that it is the components of the matter superfields which form the inflaton. 
We have then extended the model to $SO(10)$ SUSY GUTs.

Such a scenario is perfectly suited to sneutrino inflation in SUSY GUTs, allowing the inflaton to be a conjugate pair of right-handed sneutrinos, and the pair of Higgs superfields to break the GUT gauge group at the end of inflation. We have shown that in this case, if the inflaton and Higgs directions relevant for inflation lie along the right-handed neutrino direction, then this mechanism for inflation solves the monopole problem. Assuming the sneutrino trajectory for simplicity, we have then systematically examined the obvious objections to having a charged inflaton, namely the one- and two-loop gauge corrections to the potential which might be thought to threaten the flatness of the potential and so violate the slow-roll conditions, and have shown that such corrections do not pose a threat to this scheme. The key to the success of this mechanism is that the inflaton VEV during the inflationary epoch breaks the GUT gauge group but preserves D-flatness so only the F-term breaks SUSY. The inflaton therefore only couples to gauge bosons and gauginos which are heavy (and degenerate) which effectively suppresses the one- and two-loop gauge corrections. With the inclusion of SUGRA, the $\eta$-problem may be resolved by appealing to a Heisenberg symmetry which involves a modulus field stabilized during inflation.

We remark that the conjugate matter representations naturally arise from string theory constructions where generically several copies of matter of the $SO(10)$ 
$\sixteen$ and $\csixteen$, for example, appear as massless modes, where there are three more $\sixteen$s than $\csixteen$s which accounts for the three chiral families. In such a framework we are suggesting that one or more pairs of the extra $\sixteen$s and $\csixteen$s could be responsible for inflation, and their coupling to Higgs fields might trigger part of the GUT symmetry breaking at the end of inflation, without leading to excessive monopole abundance. The components of the extra $\sixteen$s and $\csixteen$s which develop VEVs during inflation lie along the right-handed neutrino directions, and mixing of these components with the physical right-handed neutrinos (in the three chiral $\sixteen$s) could lead to interesting consequences associated with reheating and non-thermal leptogenesis at the end of inflation which should be explored in future work.

In conclusion, we find that the idea that the inflaton is a gauge non-singlet is viable in the framework
of SUSY hybrid inflation,
and this opens up the possibility of having right-handed sneutrino inflation in Pati-Salam or $SO(10)$ SUSY GUTs.

\section*{Acknowledgments}
S.F.K.\ acknowledges partial support from the following grants:
STFC Rolling Grant ST/G000557/1 and a Royal Society Leverhulme Trust Senior Research
Fellowship, and also is grateful for the support and hospitality of the MPI.
S.A.\ , K.D.\ and P.M.K.\ were partially
supported by the the DFG cluster of excellence ``Origin and Structure
of the Universe''.  The work of M.B.G.\ is partially supported by the
M.E.C.\ grant FIS 2007-63364 and by the Junta de
Andaluc\'{\i}a group FQM 101.


\begin{appendix}


\section{Mass Spectrum during Inflation}\label{Mass Spectrum during Inflation}
In this appendix we calculate the masses of the relevant fields during inflation for the model of section \ref{An Example: Sneutrino Inflation}. In particular, we calculate the gauge boson masses, the fermion masses corresponding to the chiral superfields $H^c$ and $\bar H^c$ and the fermion masses arising from the mixing between the chiral and gauge multiplets. The results have been summarized in the main text in Tab.~\ref{gauge masses} and Tab.~\ref{waterfall masses} and they have been used in calculating the one loop radiative corrections in section \ref{One-Loop Corrections}. The scalar masses for the waterfall sector have been calculated in the main text, section~\ref{An Example: Sneutrino Inflation}.

\subsection{Gauge Boson Masses}\label{Gauge Boson Masses}
We now calculate the gauge boson masses corresponding to the gauge factors $SU(2)_R$ and $SU(4)_C$ of the Pati-Salam gauge group. As we will see, some of the gauge fields become massive when the inflaton fields acquire VEVs during inflation.

In our calculation, we set the coupling constants $g_R = g_C \equiv g$ close to the GUT scale and we use the following generators
\begin{equation}
\begin{split}
\mathcal{T}^a &= T_a \otimes \mathds{1}_{2 \times 2} \quad (a = 1, \, \dots \, , 15) \\
\mathcal{T}^{16} &= \mathds{1}_{4 \times 4} \otimes \tfrac{1}{2} \sigma_1 \\
\mathcal{T}^{17} &= \mathds{1}_{4 \times 4} \otimes \tfrac{1}{2} \sigma_2 \\
\mathcal{T}^{18} &= \mathds{1}_{4 \times 4} \otimes \tfrac{1}{2} \sigma_3 \; .
\end{split}
\end{equation}
Here, $\sigma_b$ are the Pauli matrices and $T_a$ are the 15 generators of $SU(4)$ displayed in Tab.~\ref{PSgenerators}.

\begin{table}[!h]
\centering
\begin{tabular}{ c c c}
\vspace{0.3cm}
$T_1=\frac{1}{2}
\begin{pmatrix}
0&1&0&0\\
1&0&0&0\\
0&0&0&0\\
0&0&0&0
\end{pmatrix}$ &
$T_2=\frac{1}{2}
\begin{pmatrix}
0&-\imag&0&0\\
\imag&0&0&0\\
0&0&0&0\\
0&0&0&0
\end{pmatrix}$ &
$T_3=\frac{1}{2}
\begin{pmatrix}
1&0&0&0\\
0&-1&0&0\\
0&0&0&0\\
0&0&0&0
\end{pmatrix}$\\
\vspace{0.3cm}
$T_4=\frac{1}{2}
\begin{pmatrix}
0&0&1&0\\
0&0&0&0\\
1&0&0&0\\
0&0&0&0
\end{pmatrix}$ &
$T_5=\frac{1}{2}
\begin{pmatrix}
0&0&-\imag&0\\
0&0&0&0\\
\imag&0&0&0\\
0&0&0&0
\end{pmatrix}$ &
$T_6=\frac{1}{2}
\begin{pmatrix}
0&0&0&0\\
0&0&1&0\\
0&1&0&0\\
0&0&0&0
\end{pmatrix}$\\
\vspace{0.3cm}
$T_7=\frac{1}{2}
\begin{pmatrix}
0&0&0&0\\
0&0&-\imag&0\\
0&\imag&0&0\\
0&0&0&0
\end{pmatrix}$ &
$T_8=\frac{1}{2\sqrt{3}}
\begin{pmatrix}
1&0&0&0\\
0&1&0&0\\
0&0&-2&0\\
0&0&0&0
\end{pmatrix}$ &
$T_9=\frac{1}{2}
\begin{pmatrix}
0&0&0&1\\
0&0&0&0\\
0&0&0&0\\
1&0&0&0
\end{pmatrix}$\\
\vspace{0.3cm}
$T_{10}=\frac{1}{2}
\begin{pmatrix}
0&0&0&-\imag\\
0&0&0&0\\
0&0&0&0\\
\imag&0&0&0
\end{pmatrix}$ &
$T_{11}=\frac{1}{2}
\begin{pmatrix}
0&0&0&0\\
0&0&0&1\\
0&0&0&0\\
0&1&0&0
\end{pmatrix}$ &
$T_{12}=\frac{1}{2}
\begin{pmatrix}
0&0&0&0\\
0&0&0&-\imag\\
0&0&0&0\\
0&\imag&0&0
\end{pmatrix}$\\
\vspace{0.3cm}
$T_{13}=\frac{1}{2}
\begin{pmatrix}
0&0&0&0\\
0&0&0&0\\
0&0&0&1\\
0&0&1&0
\end{pmatrix}$ &
$T_{14}=\frac{1}{2}
\begin{pmatrix}
0&0&0&0\\
0&0&0&0\\
0&0&0&-\imag\\
0&0&\imag&0
\end{pmatrix}$ &
$T_{15}=\frac{1}{2\sqrt{6}}
\begin{pmatrix}
1&0&0&0\\
0&1&0&0\\
0&0&1&0\\
0&0&0&-3
\end{pmatrix}$
\end{tabular}
\caption{\label{PSgenerators} Fifteen $SU(4)_C$ generators.}
\end{table}

The masses for the gauge bosons are given by the following term in the Lagrangian
\begin{align}\label{GaugeBosonMasses}
\mathcal{L}_{GB}
&= \left| \sum_{a = 1}^{18} g \, A^a_{\mu} \, \mathcal{T}^a \langle R^c \rangle \right|^2 + \; \text{terms for} \; \langle \bar R^c \rangle \; ,
\end{align}
where $\langle R^c \rangle \, , \, \langle \bar R^c \rangle$ are the VEVs of the sneutrinos acting as inflatons, cf.~Eq.~\eqref{infvevs-1}.

We can easily see that the gauge fields corresponding to the generators $\mathcal{T}^1, \, \dots \, , \mathcal{T}^8$ remain massless.
On the other hand, for the gauge fields corresponding to the generators $\mathcal{T}^9$ and $\mathcal{T}^{10}$ we find
\begin{align}
\mathcal{L}_{GB}  \supset  \; \frac{1}{2} \, g^2 \,{\langle \nu^c \rangle}^2 \left[ (A_\mu^9)^2 + (A_\mu^{10})^2 \right].
\end{align}
This yields
\begin{equation}
m_9^2 = m_{10}^2 = g^2 \,{\langle \nu^c \rangle}^2.
\end{equation}
Similarly, the gauge bosons corresponding to the generators $\mathcal{T}^{11}, \, \dots \, , \mathcal{T}^{14}$ as well as $\mathcal{T}^{16}$ and  $\mathcal{T}^{17}$ acquire the same mass.
The generators $\mathcal{T}^{18}$ and $\mathcal{T}^{15}$ are diagonal and the corresponding gauge bosons mix. We find
\begin{align}\label{Masses1518}
\mathcal{L}_{GB} \supset  \; g^2 \frac{{\langle \nu^c \rangle}^2}{4} \left(\, A_\mu^{18} - \, \sqrt{3\over 2} \, A_\mu^{15}\right)^2
+ \;\text{terms for} \;\langle \bar R^c \rangle \; .
\end{align}
Defining the new normalized field
\begin{equation}
Z^{\scriptscriptstyle{\parallel}}_\mu \equiv \sqrt {\frac{2}{5}} \left(A^{18}_\mu - \sqrt{\frac{3}{2}} A^{15}_\mu \right)
\end{equation}
this becomes
\begin{align}
\mathcal{L}_{GB} \supset  \frac{5}{4} g^2 \, {\langle \nu^c \rangle}^2 \,  (Z^{\scriptscriptstyle{\parallel}}_\mu)^2 .
\end{align}

\noindent
The combination orthogonal to $Z^{\scriptscriptstyle{\parallel}}_\mu$, i.e
\begin{equation}
Z^{\scriptscriptstyle{\perp}}_\mu \equiv \sqrt {\frac{2}{5}}  \left(A^{15}_\mu + \sqrt{\frac{3}{2}} A^{18}_\mu \right)
\end{equation}
remains massless. The gauge boson masses have been summarized in Tab.~\ref{gauge masses}.

\subsection{Fermion Mass Spectrum}

In a SUSY theory there are two contributions to the fermion masses, one coming directly from the superpotential and another one from the mixing between the chiral and the gauge multiplets.

The contribution from the superpotential is given by
\begin{equation}
\mathcal{L}_1
=- \frac{1}{2} \,\frac{\delta^2 W}{\delta\phi_i \,\delta \phi_j} \left(\psi_i \,\psi_j + \bar\psi_i \,\bar\psi_j\right) \, . \label{Fermion mass}
\end{equation}
Here, $\phi_i$ and $\psi_i$ are the scalar boson and chiral fermion contained in the chiral superfield \mbox{$\Phi_i \ni \phi_i , \psi_i$} and $W$ is the superpotential regarded as a function of the scalar fields only.

Using the form of the superpotential in Eq.~(\ref{super4c}) and keeping in mind that the VEVs of the scalar components of $H^c$ and $\bar H^c$ remain at zero during inflation, we conclude that Eq.~(\ref{Fermion mass}) does not contribute to the fermion masses corresponding to the chiral multiplets $R^c$ and $\bar R^c$. But it does contribute to the fermion masses corresponding to $H^c$ and $\bar H^c$:
\begin{align}
\mathcal{L}_1 = & - \zeta \,{\langle \nu^c \rangle}^2
\left[ \,\psi_{u^c_{1H}} \psi_{\ou^c_{1H}} + \, \dots \, + \psi_{d^c_{3H}} \psi_{\od^c_{3H}} + \psi_{e^c_H} \psi_{\oel^c_H} + \,\text{h.c.}
\right]\nonumber \\
& - \frac{1}{2} \,{\langle \nu^c \rangle}^2 \left[ \,2\,\gamma\, \psi_{\nu^c_H}\psi_{\nu^c_H} + 2\,(\zeta + \xi) \,\psi_{\nu^c_H} \psi_{\on^c_{H}} +
2\,\lambda\, \psi_{\on^c_{H}}\psi_{\on^c_{H}} +\, \text{h.c.} \right] \,.
\end{align}
Combining two chiral spinors to a Dirac spinor
\begin{equation}
\Psi_{u^c_{1H}} =
\begin{pmatrix}
\psi_{u^c_{1H}} \\
\bar\psi_{\ou^c_{1H}}
\end{pmatrix} \;\, , \quad \dots
\end{equation}
the first part becomes
\begin{equation}
\mathcal{L}_1 \supset - \zeta\, {\langle \nu^c \rangle}^2
\left[ \,\bar\Psi_{u^c_{1H}} \Psi_{u^c_{1H}} + \, \dots \, + \bar\Psi_{d^c_{3H}} \Psi_{d^c_{3H}} + \bar\Psi_{e^c_H} \Psi_{e^c_{H}} \right]\,.
\end{equation}
Diagonalizing the mass matrix of the second part, we find
\begin{equation}
\mathcal{L}_1 \supset - \frac{1}{2} \,{\langle \nu^c \rangle}^2
\left[\, (2 \,\gamma - \zeta - \xi)\, \psi_a \,\psi_a \,+\, (2 \,\gamma + \zeta + \xi) \, \psi_b \,\psi_b +\, \text{h.c.}  \right]\,,
\end{equation}
where
\begin{equation}
\begin{pmatrix}
\psi_a \\
\psi_b
\end{pmatrix} =
\frac{1}{\sqrt{2}}\,
\begin{pmatrix}
\psi_{\on^c_{H}} - \,\psi_{\nu^c_H} \\
\psi_{\on^c_{H}} + \,\psi_{\nu^c_H}
\end{pmatrix}\,,
\end{equation}
and we have set $\gamma = \lambda$ for simplicity.

Finally, defining the two Majorana spinors
\begin{equation}
\Psi_a =
\begin{pmatrix}
\psi_a\\
\bar\psi_a
\end{pmatrix}\,, \qquad
\Psi_b =
\begin{pmatrix}
\psi_b \\
\bar\psi_b
\end{pmatrix}
\end{equation}
this becomes
\begin{equation}
\mathcal{L}_1 \supset - \frac{1}{2} \,{\langle \nu^c \rangle}^2
\left[\,\left(2 \,\gamma - \zeta - \xi\right) \, \bar\Psi_a \Psi_a \,+\, \left(2 \,\gamma + \zeta + \xi\right) \, \bar\Psi_b \Psi_b \right]\,.
\end{equation}
The resulting masses have been summarized in Table \ref{waterfall masses}.

Next, we turn to the second contribution due to the mixings between the chiral fermions $\psi_i$ of the chiral superfields and the gauginos. It is given by
\begin{equation}
\mathcal{L}_2 = - \sqrt{2} g \, \sum_a \left(\phi^{*}_{R^c} \, \mathcal{T}^a \, \psi_{R^c} \right) \, \lambda^a
- \sqrt{2} g \, \sum_a \, \bar\lambda^{a} \left(\bar\psi_{R^c} \, \mathcal{T}^a \, \phi_{R^c} \right) \, + \text{terms for} \bar R^c,
\end{equation}
where $\phi_{R^c}$ and $\psi_{R^c}$ are the scalar and fermionic fields contained in the chiral supermultiplet $R^c$.

Plugging in the VEVs of the $R^c$ and $\bar R^c$ fields
we end up with
\begin{equation}
\begin{split}
\mathcal{L}_2 =  - \frac{g}{\sqrt{2}} \,\langle \nu^c \rangle \,
\Big[ \, &\psi_{u^c_1} \left(- \lambda^9 + \,\text{i} \,\lambda^{10}\right)
\,+\, \psi_{\ou^c_1} \left(\lambda^9 + \,\text{i} \,\lambda^{10}\right)+ \; \dots \; +\\[-0.23cm]
&\psi_{u^c_3} \left(- \lambda^{13} + \,\text{i} \,\lambda^{14}\right)
\,+\, \psi_{\ou^c_3} \left(\lambda^{13} + \,\text{i} \,\lambda^{14}\right)+ \\
&\psi_{e^c} \left(- \lambda^{16} - \,\text{i} \,\lambda^{17}\right)
\,+\, \psi_{\oel^c} \left(\lambda^{16} - \,\text{i} \,\lambda^{17}\right) +\\[-0.1cm]
&\psi_{\nu^c} \left(\sqrt{\tfrac{3}{2}} \,\lambda^{15} - \lambda^{18}\right)
\,+\, \psi_{\on^c} \left(- \sqrt{\tfrac{3}{2}}\,\lambda^{15} + \lambda^{18}\right)+ \,\text{h.c.} \Big] \, .
\end{split}
\end{equation}
Defining the following normalized left-chiral fields
\begin{align}
\chi_1 &= \tfrac{1}{\sqrt{2}} \left(- \lambda^9 + \,\text{i}\, \lambda^{10}\right) &
\chi_2 &= \tfrac{1}{\sqrt{2}} \left(\lambda^9 + \,\text{i}\, \lambda^{10}\right) \nonumber \\
& \hspace{5.2cm} \dots & & \nonumber \\
\chi_{e^c} &= -\tfrac{1}{\sqrt{2}} \left(\lambda^{16} + \,\text{i}\, \lambda^{17}\right) &
\chi_{\oel^c} &= \tfrac{1}{\sqrt{2}} \left(\lambda^{16} - \,\text{i}\, \lambda^{17}\right) \\
\psi_\nu^{\scriptscriptstyle{\parallel}} &= \tfrac{1}{\sqrt{2}} (\psi_{\nu^c} - \psi_{\bar\nu^c}) &
\psi_\nu^{\scriptscriptstyle{\perp}} &= \tfrac{1}{\sqrt{2}} (\psi_{\nu^c} + \psi_{\bar\nu^c}) \nonumber \\
\chi_{\nu^c}^{\scriptscriptstyle{\parallel}} &= \sqrt{\tfrac{2}{5}} \left(\sqrt{\tfrac{3}{2}} \lambda^{15} - \lambda^{18}\right) &
\chi_{\nu^c}^{\scriptscriptstyle{\perp}} &= \sqrt{\tfrac{2}{5}} \left(\sqrt{\tfrac{3}{2}} \lambda^{18} + \lambda^{15}\right) \nonumber
\end{align}
we can combine these with the chiral fermion fields from the $R^c$ and $\oR^c$ superfields to form the following Dirac spinors
\begin{equation}
\Psi_1 =
\begin{pmatrix}
\psi_{u^c_1} \\
\bar \chi_1
\end{pmatrix} \quad
\Psi_2 =
\begin{pmatrix}
\psi_{\bar u^c_1} \\
\bar \chi_2
\end{pmatrix} \hspace{0.5cm} \dots \hspace{0.5cm}
\Psi_{\nu^c}^{\scriptscriptstyle{\parallel}} =
\begin{pmatrix}
\psi_{\nu^c}^{\scriptscriptstyle{\parallel}} \\
\bar \chi_{\nu^c}^{\scriptscriptstyle{\parallel}}
\end{pmatrix} \quad
\Psi_{\nu^c}^\perp =
\begin{pmatrix}
\psi_{\nu^c}^\perp \\
\bar \chi_{\nu^c}^\perp
\end{pmatrix}.
\end{equation}
With these, we can now write
\begin{equation}
\mathcal{L}_2 = - \,g \,\langle \nu^c \rangle  \left[\bar\Psi_1 \Psi_1 \,+ \, \ldots \,+\, \bar\Psi_6 \Psi_6
\,+\, \bar\Psi_{e^c} \Psi_{e^c} \,+\, \bar\Psi_{\oel^c} \,\Psi_{\oel^c} \right]
- \sqrt{\frac{5}{2}} \,g \,\langle \nu^c \rangle \, \bar\Psi_{\nu^c}^{\scriptscriptstyle{\parallel}} \Psi_{\nu^c}^{\scriptscriptstyle{\parallel}}.
\end{equation}
The mass spectrum has been listed in Tab.~\ref{gauge masses}.


\section{Effective Dimension 5 Operators in Pati-Salam}\label{Effective Dimension 5 Operators in Pati-Salam}
In our simple Pati-Salam model of section \ref{The Model} we want to consider all effective dimension 5 operators which are generated by the exchange of singlet messenger fields and are allowed by the imposed $R$ and $\Zten$ symmetries.

\begin{figure}[!h]
\psfrag{R}{$R^c$}
\psfrag{Hb}{$\oH^c$}
\psfrag{D1}{$\Delta_1$}
\psfrag{Rb}{$\oR^c$}
\psfrag{H}{$H^c$}
\psfrag{D2}{$\Delta_2$}
\psfrag{D3}{$\Delta_3$}
\psfrag{D4}{$\Delta_4$}
\center
\includegraphics{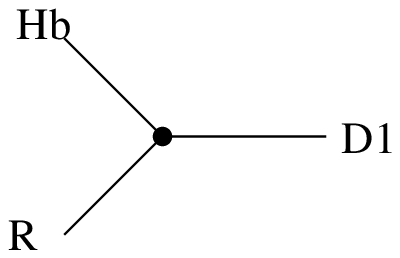}
\hspace{1cm}
\includegraphics{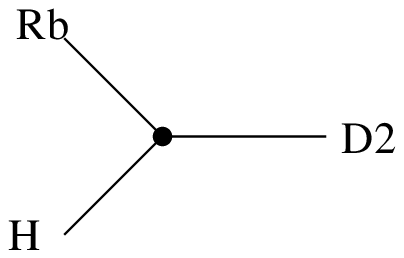}\\
\vspace{1cm}
\includegraphics{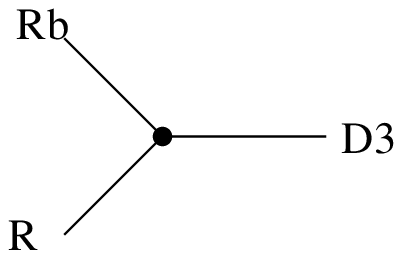}
\hspace{1cm}
\includegraphics{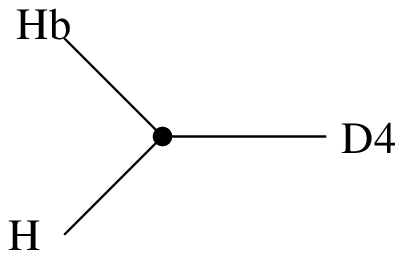}
\caption{\label{FundamentalVertices} Interaction vertices yielding singlet messenger fields.}
\end{figure}

To begin with, let us focus on the $SU(4)_C$ gauge structure. Under $SU(4)_C$ we have $\oR^c , \oH^c \sim \rep{4}$, whereas
$R^c , H^c \sim \bar{\rep{4}}$. We know that
\begin{equation}
\begin{split}
\rep{4} \otimes \bar{\rep{4}} = \rep{1} \oplus \rep{15} \\
\rep{4} \otimes \rep{4} = \rep{10} \oplus \bar{\rep{6}} \\
\bar{\rep{4}} \otimes \bar{\rep{4}} = \bar{\rep{10}} \oplus \rep{6}
\end{split}
\end{equation}
To form a singlet messenger we therefore have to couple one field transforming as a $\rep{4}$ to one transforming as a $\bar{\rep{4}}$. (Coupling two such fields will also yield a singlet under $SU(2)_R$, since in our model they transform as $\rep{2}$ respectively $\bar{\rep{2}}$ under this symmetry.) The allowed fundamental vertices are shown in figure \ref{FundamentalVertices}.

\begin{figure}
\psfrag{D1}{$\Delta_i$}
\psfrag{D2}{$\Delta_j$}
\center
\includegraphics{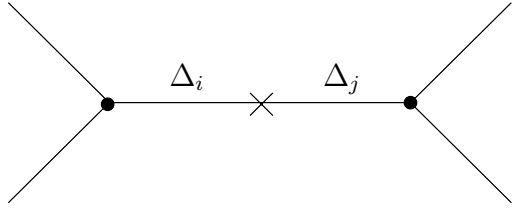}
\caption{\label{MassInsertion} Feynman diagram generating the effective $d=5$ operators.}
\end{figure}

When combining two of these fundamental vertices to form an effective $d=5$ operator, we have to introduce a mass insertion into the diagram, cf.~figure \ref{MassInsertion}.
The corresponding term in the superpotential reads
\begin{equation}
W \supset \Lambda \, \Delta_i \Delta_j \;.
\end{equation}
From this we see that the $R$ and $\Zten$ quantum numbers of the messenger fields involved have to add up to $1$ respectively a multiple of $10$. These quantum numbers can be found in Tab.~\ref{QuantumNumbersMessenger}

\begin{table}[!h]
\center
\begin{tabular}{c c c}\hline
Messenger & $R$ & $\Zten$ \\ \hline
$\Delta_1$ & 1/2 & 5 \\
$\Delta_2$ & 1/2 & 5\\
$\Delta_3$ & 0 & 3\\
$\Delta_4$ & 1 & 7\\
\hline
\end{tabular}
\caption{\label{QuantumNumbersMessenger} Quantum numbers of the singlet messenger fields.}
\end{table}

Thus, we can couple $\Delta_1$ and $\Delta_2$ to themselves, $\Delta_1$ to $\Delta_2$ and finally $\Delta_3$ to $\Delta_4$. After integrating out the heavy messengers, the following effective operators are generated, where round brackets denote contraction of the $SU(4)_C$ and $SU(2)_R$ indices
\begin{equation}
\begin{split}
\mathcal{O}_1^{d=5} &= \frac{\lambda}{\Lambda}\,\left(R^c \oH^c\right)\left(R^c \oH^c\right) \\
\mathcal{O}_2^{d=5} &= \frac{\gamma}{\Lambda}\,\left(\oR^c H^c\right)\left(\oR^c H^c\right) \\
\mathcal{O}_3^{d=5} &= \frac{\zeta}{\Lambda}\,\left(R^c \oR^c\right)\left(H^c \oH^c\right) \\
\mathcal{O}_4^{d=5} &= \frac{\xi}{\Lambda}\,\left(R^c \oH^c\right)\left(\oR^c H^c\right)
\end{split}
\end{equation}

\begin{figure}[!h]
\psfrag{R}{$R^c$}
\psfrag{Hb}{$\oH^c$}
\psfrag{Rb}{$\oR^c$}
\psfrag{H}{$H^c$}
\psfrag{L}{$\lambda_{ij}$}
\psfrag{G}{$\gamma$}
\psfrag{Z}{$\zeta_i$,$\xi_i$}
\center
\includegraphics{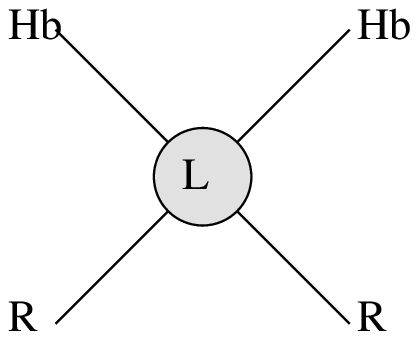}
\hspace{0.5cm}
\includegraphics{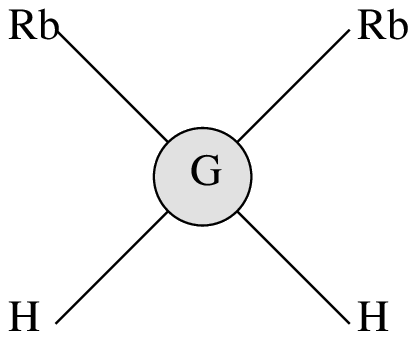}
\hspace{0.5cm}
\includegraphics{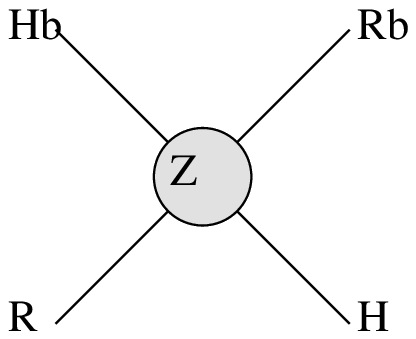}
\caption{\label{Effectived5Operators} Generated effective $d=5$ operators.}
\end{figure}

The complete (effective) superpotential now reads
\begin{equation}\label{PatiSalamSuperpotential}
\begin{split}
W = &\,\kappa \,S\left( \frac{\langle X\rangle}{\Lambda}\,H^c \oH^c - M^2\right) \\
&+ \frac{\lambda}{\Lambda}\,(R^c \oH^c)(R^c \oH^c)
+ \frac{\gamma}{\Lambda} \,(\oR^c H^c)(\oR^c H^c)
+ \frac{\zeta}{\Lambda}\,(R^c \oR^c)(H^c \oH^c)
+ \frac{\xi}{\Lambda}\,(R^c \oH^c)(\oR^c H^c) \,.
\end{split}
\end{equation}

\end{appendix}


\end{document}